\documentclass[reqno]{amsart}
\usepackage{amsmath}
\usepackage{amssymb}
\usepackage{amsfonts}
\usepackage{amsthm}
\usepackage{amstext}
\usepackage{enumerate}
\usepackage[english]{babel}

     \newcommand{\supp}{\operatorname{supp}}
     \newcommand{\Spec}{\operatorname{Spec}}
     \newcommand{\Div}{{\operatorname{div}}}
     \newcommand{\tr}{{\operatorname{tr}}}

     \newcommand{\curl}{{\operatorname{curl}\,}}
     \newcommand{\Ran}{{\operatorname{Ran}\,}}

     \theoremstyle{plain}
     \newtheorem{thm}{Theorem}[section]
     \newtheorem{prop}[thm]{Proposition}
     \newtheorem{lemma}[thm]{Lemma}
     \newtheorem{cor}[thm]{Corollary}
     
     \theoremstyle{definition}

     \newtheorem{remark}[thm]{Remark}

     \numberwithin{equation}{section}

\title{Spectral confinement and current for atoms in strong magnetic fields}
\author{S. Fournais}
\address{S. Fournais\\ CNRS and 
Laboratoire de Math\'{e}matiques\\
Universit\'{e} Paris-Sud - B\^{a}t 425\\
F-91405 Orsay Cedex\\ France.}

\thanks{The author is partly supported by the European Research Network
`Postdoctoral Training Program in Mathematical Analysis of Large
Quantum Systems' with contract number HPRN-CT-2002-00277, and the ESF
Scientific Programme in Spectral Theory and Partial Differential
Equations (SPECT)}

\email{soeren.fournais@math.u-psud.fr}
\date{26/8/2006}

\begin{document}

\begin{abstract}
We study confinement of the ground state of atoms in strong magnetic fields to different subspaces related to the lowest Landau band. The results obtained allow us to calculate the quantum current in the entire semiclassical region $B \ll Z^3$.
\end{abstract}

\maketitle

\thispagestyle{empty}

\tableofcontents

\section{Introduction and main results}
\subsection{Introduction}\label{Intro}~\\
Large atoms (e.g.~iron) subject to strong magnetic fields exist in nature on the surface of neutron stars. A large amount of research in physics and mathematics has been devoted to the study of this system. Of particular importance for the present work are the articles by Lieb, Solovej and Yngvason \cite{LSY1,LSY} (see also \cite{Yngvason}) which can be seen as the mathematical starting point of the investigation of the limits and approximating models considered in this paper and contain a large number of references to earlier work in the physics literature.

We describe the atoms by non-relativistic quantum mechanics in the fixed-nucleus approximation.
Remembering the spin of the electrons, the basic operator for an atom in a
magnetic field is therefore the Pauli Hamiltonian 
\begin{align}
   \label{eq:hamiltonian}
   H(N,Z,{\bf A}) = &\sum_{j=1}^N \left( H_{{\bf A}}^{(j)}
-\frac{Z}{|x^{(j)}|} \right)
   + \sum_{1\leq j < k \leq N} \frac{1}{|x^{(j)}-x^{(k)}|}.
\end{align}
Here $H_{{\bf A}}=({\bf p}+ {\bf A}(x))^2 + {\bf \sigma} \cdot {\bf
B}(x)$, 
with ${\bf B} = \curl {\bf A}$, and
${\bf \sigma}=(\sigma_1,\sigma_2,\sigma_3)$ is the
vector of
Pauli spin matrices,
\begin{align*}
   \sigma_1 = \left(\begin{array}{cc} 0&1\\1&0 \end{array} \right),
\,\,\,\,\,\,\,\,\,\,\,
   \sigma_2 = \left(\begin{array}{cc} 0&-i\\i&0 \end{array} \right),
\,\,\,\,\,\,\,\,\,\,\,
   \sigma_3 = \left(\begin{array}{cc} 1&0\\0&-1 \end{array} \right) .
\end{align*}
The operator $H(N,Z,{\bf A})$
acts on the electronic Hilbert space (including spin)
${\mathcal H}=\wedge_{j=1}^N L^2({\mathbb R}^3; {\mathbb
C}^2)$, ${\bf p} =(p_1,p_2,p_3) = (-i\nabla)$ and a
superscript $ (j) $ denotes that the corresponding operator
acts on the $j$-th factor in the
product. In particular $x^{(j)}$ is the coordinate of the
$j$-th electron. We will also use the notation
${\bf p}_{{\bf A}} = (p_{{\bf A},1},p_{{\bf A}, 2},p_{{\bf
A},3}) = ({\bf p}+ {\bf A})$. The scalar product in ${\mathcal H}$ is
denoted by $\langle \cdot, \cdot
\rangle$.

We consider the case of constant magnetic field, i.e. we fix ${\bf B} =
(0,0,B)$ and ${\bf A}(x) =
\tfrac{1}{2} {\bf B} \times x$. The magnetic field will be {\it
strong}, which mathematically means that we study the limit $B
\rightarrow +\infty$. At the same time the atoms will be {\it large},
which informally means that $N=Z$ and that $Z \rightarrow \infty$ (the
restriction $N=Z$ is slightly too strong, see Theorem~\ref{thm:loc_improved} for the actual
assumption). The limiting behaviour depends on the relative size of $Z$ and $B$.

In the constant field case, i.e. with ${\bf B}, {\bf A}$ as above, we write $H(N,Z,B)$ instead of
$H(N,Z,{\bf A})$, and define the ground state {\it energy} of the atom
by
\begin{eqnarray}
   \label{eq:7}
   E(N,Z,B) = \inf \Spec H(N,Z,B).
\end{eqnarray}
We sometimes denote the quantum energy $E(N,Z,B)$ by $E^{\rm Q}(N,Z,B)$ in order to distinguish it from other energies appearing in the paper. It is known that for $N \leq Z+1$, and all $B>0$, the ground state energy $E(N,Z,B)$ is a discrete eigenvalue below the essential spectrum of $H(N,Z,B)$ (see \cite{AHS3}).

Heuristically one can get the right order of magnitude of the energy from the following description (taken from \cite{LSY-conference})\footnote{We use the intuitive notations $\ll, \approx; \lesssim$, in discussions, results will be more precisely stated. For instance, the statement $B \ll Z^{4/3}$ means that we consider sequences $\{(B_n, Z_n)\}_{n \in {\mathbb N}}$ such that $B_n Z_n^{-4/3} \rightarrow 0$ (together with the standing assumption that $Z_n \rightarrow \infty$) as $n \rightarrow \infty$. }.

\begin{itemize}
\item ${\mathbf{ B \ll Z^{4/3}}}$.\\
For small $B$, we can think of each electron as occupying a spherical region of space of radius $a$. The kinetic energy per particle is therefore of order $a^{-2}$. The small spheres organize to form a large sphere of radius $R$, in order to minimize the electrostatic energy which becomes of order $Z/R$. Using the volume relation $R^3 \approx N a^3$ and setting the kinetic energy equal to the potential energy, we find 
\begin{align*}
a \sim Z^{-2/3}, \quad\quad R\sim Z^{-1/3}, \quad\quad E \sim -Z^{7/3}.
\end{align*}
The last expression, $E \sim -Z^{7/3}$, is the order of magnitude of the energy in standard, non-magnetic, Thomas-Fermi theory. Notice that the magnetic field did not enter in the discussion.\\
In this region standard Thomas-Fermi theory correctly describes the ground state energy to leading order.
\item ${\mathbf{ Z^{4/3} \lesssim B \ll Z^3}}$.\\
For large $B$, the magnetic length scale $B^{-1/2}$ becomes smaller than the radius $Z^{-2/3}$ of the electron `sphere'. The shape of the `electron' now becomes that of a cylinder, with axis parallel to the magnetic field, of radius $B^{-1/2}$ and length $L$.
The kinetic energy per particle is $L^{-2}$, since the Pauli kinetic energy vanishes in the perpendicular variables. 
The electronic cylinders organize in a sphere of radius $R$.
Proceeding as before, we find
\begin{align*}
L \sim Z^{-2/3} B^{-1/5}, \quad\quad R\sim Z^{1/5} B^{-2/5}, \quad\quad E \sim  -Z^{9/5}B^{2/5}.
\end{align*}
In this region a magnetic Thomas-Fermi theory (MTF) correctly describes the ground state energy of the atom to leading order. For $B \gg Z^{4/3}$ this MTF-theory simplifies since only the lowest Landau band has to be taken into account.
\item ${\mathbf{ Z^3 \lesssim B}}$.\\
When $B$ is above $Z^3$ the length of each individual cylinder becomes comparable to the radius of the atom and a spherical arrangement ceases to be possible. The atom as such becomes cylindrical with radius $R$ and length $L$ with
\begin{align*}
L \sim Z^{-1}[ \log(B/Z^3) ]^{-1}, \quad\quad R\sim \sqrt{Z/B}, \quad\quad E \sim -Z^3 [ \log(B/Z^3) ]^{2}.
\end{align*}
In \cite{LSY1} a density matrix functional was introduced and analysed and it was shown that it correctly predicts the ground state energy of the atom for $B \gg Z^{4/3}$ in particular it is valid for $Z^3 \lesssim B$.
\end{itemize}

For later notational convenience we define a function ${\mathcal E} = {\mathcal E}(Z,B)$ that gives the magnitude of the ground state energy.
\begin{align}
\label{eq:CalE}
{\mathcal E}(Z,B) := \begin{cases}
Z^{7/3}, & B \leq Z^{4/3},\\
B^{2/5} Z^{9/5}, &Z^{4/3}\leq B \leq 2Z^3,\\
Z^3 (\log \frac{B}{Z^3})^2, & 2Z^3 < B.
\end{cases}
\end{align}
It was proven by Lieb, Solovej and Yngvason that, for all $\lambda>0$, there exists $C>0$ such that,
if $\lambda = N/Z $ and $Z\geq C$, then
\begin{align}
C^{-1}{\mathcal E}(Z,B)\leq|E^Q(N,Z,B)| \leq C{\mathcal E}(Z,B).
\end{align}
Actually they prove much more, in particular, they introduce a number of approximating functionals, depending on the asymptotic regions above, and prove convergence of the ground state energy of the quantum model to that of the approximating functional. In the regime(s) where $B \ll Z^3$ one possibility for the approximating model is a Thomas-Fermi type theory, depending on the magnetic field, which simplifies in the limit $Z^{4/3} \ll B$ since only the lowest Landau band needs to be taken into account.
Magnetic Thomas-Fermi theory will be discussed in subsection~\ref{ResCurrent} below and in Section~\ref{CurrentMTF}.
For work on these questions see \cite{LSY1, LSY}, \cite{HS2000} and \cite{ES,ES1,ES3,ES4} for the case of non-constant magnetic field.

It is of interest to determine how well these approximating theories reproduce the results of the full quantum model. Apart from the known approximation of the ground state energy, one can ask whether the ground state {\it density} and {\it current} are correctly predicted.  
The density $\rho \in L^1({\mathbb R}^3)$ of the wavefunction $\psi \in \wedge_{j=1}^N L^2({\mathbb R}^3,{\mathbb C}^2)$ is the function\footnote{Remember that $\psi$ takes values in $\otimes_{j=1}^N {\mathbb C}^2= {\mathbb C}^{2^N}$, so the norm, $|\cdot |$, in the expression for $\rho$, is the Euclidean norm in $ {\mathbb C}^{2^N}$.}
$$
\rho(x) = N \int_{{\mathbb R}^{3N-3}} |\psi(x,x_2,\ldots, x_N)|^2\,dx_2\cdots dx_N.
$$
The current ${\bf j}$ of $\psi$ takes values in ${\mathbb R}^3$ and is most conveniently expressed in the following weak sense. For all ${\bf a} \in C_0^{\infty}({\mathbb R}^3, {\mathbb R}^3)$ we define
\begin{align}
\label{eq:CurrDef}
\int_{{\mathbb R}^3} {\bf a} \cdot {\bf j}\,dx :=
\langle \psi, J({\bf a}) \psi \rangle,
\end{align}
where, with ${\bf b} = (b_1,b_2,b_3) = \curl {\bf a}$,
\begin{align}
\label{eq:CurrOp}
J({\bf a}) = B\sum_{j=1}^N \Big\{ {\bf a}(x^{(j)}) \cdot {\bf p}_{{\bf A}}^{(j)}+{\bf p}_{{\bf A}}^{(j)}\cdot {\bf a}(x^{(j)})
+
{\bf b}(x^{(j)}) \cdot \sigma^{(j)} \Big\}.
\end{align}

The leading order behaviour of the density was already calculated in \cite{LSY1,LSY} and it was seen that all the approximating models give the correct asymptotics in their respective regimes of validity for the energy. The calculation of the current is harder. In \cite{Fournais3} it was proven that a sequence of approximating ground states $\psi_{Z,B}$, i.e.~a sequence of normalised functions with
$$
\langle \psi_{Z,B},  H(N,Z,B) \psi_{Z,B} \rangle - E^Q(N,Z,B) = o(E^Q(N,Z,B)),
$$ 
does not necessarily give the correct current. 
However, the main result of the present paper is that MTF correctly predicts the leading order term of the current in its entire regime of validity. This is stated more precisely as Theorem~\ref{thm:summarize_improved} below.

Before continuing the discussion of the current let us recall that the {\it magnetisa\-tion} ${\bf M}$ of the atom is related to the current by the relation
$$
\curl {\bf M} = {\bf j}.
$$
We also recall that the simpler question of calculating the {\it total magnetisation} has been answered in \cite{fournais6}.

\subsection{Results on the current}
\label{ResCurrent}~\\
Let $T$ be the symmetry of reflection in the plane perpendicular to the magnetic field: 
\begin{align}
(T\psi)\big(x^{(1)},\ldots,x^{(N)}\big)
=\psi\left(\overline{x^{(1)}},\ldots,\overline{x^{(N)}}\right),
\end{align}
with
$\overline{x} = \overline{(x_1,x_2,x_3)} = (x_1,x_2,-x_3)$. 
Clearly, the symmetry $T$ commutes with $H(N,Z,B)$.

We now introduce the MTF-functional (see
\cite{LSY1, LSY, fournais8} for further details on ${\rm MTF}$-theory). To a density $\rho$ and a magnetic field ${\bf B}$ one can associate an energy ${\mathcal E}_{Z, {\bf B}}^{\rm MTF}[\rho]$ by
\begin{align}
\label{eq:E-MTF}
{\mathcal E}_{Z, {\bf B}}^{\rm MTF}[\rho] =
\int_{{\mathbb R}^3} \Big(\tau_{|{\bf B}(x)|}^{\rm MTF}(\rho(x)) - Z \frac{\rho(x)}{|x|}\Big)\,dx + D(\rho,\rho),
\end{align}
where $D(f,g)$ denotes the direct Coulomb energy,
$$
D(f,g) := \frac{1}{2}\int \frac{\overline{f(x)}g(y)}{|x-y|} \,dxdy,
$$
and the `kinetic energy', $\tau_b^{\rm MTF}(\rho)$, is defined by
$$
\tau_b^{\rm MTF}(\rho):=\sup_{v \geq 0} \big( \rho v - P_b(v) \big),
$$
with
\begin{align}
\label{eq:Pb}
P_b(v) := \frac{b}{3\pi^2} \Big( v^{3/2} + 2 \sum_{j=1}^{\infty} [2jb -v]_{-}^{3/2}\Big).
\end{align}
In the expression for $P_b$, $[\,\cdot\,]_{-}$ denotes the negative part: $[x]_{-} := \max\{ 0, -x \}$. 
Notice that we can take a non-constant magnetic field in the definition of the MTF-functional.

We denote by
$E^{{\rm MTF}}(N,Z, {\bf B})$ the atomic ground state energy in
magnetic Tho\-mas-Fermi theory---with $N$ electrons, nuclear charge $Z$ and
in the presence of the external magnetic field ${\bf B}$---defined as
\begin{align}
E^{{\rm MTF}}(N,Z, {\bf B}) := \inf_{\rho \in {\mathcal C}_{N,{\bf B}} }
{\mathcal E}_{Z, {\bf B}}^{\rm MTF}[\rho]  , 
\end{align}
with 
\begin{align}
\label{eq:DefCB}
{\mathcal C}_{N,{\bf B}} := \Big\{ \rho \in L^1({\mathbb R}^3) \,\Big|\, & 0 \leq \rho \, ,
\, \int_{{\mathbb R}^3} \rho(x) \,dx \leq N, \, \nonumber\\
&D(\rho,\rho) < \infty \quad \text{ and } \quad \int_{{\mathbb R}^3} \tau_{|{\bf B}(x)|}^{\rm MTF}(\rho(x))\,dx < \infty
\Big\}.
\end{align}

Let furthermore ${\bf e}_z=(0,0,1)$ be the third standard unit vector in ${\mathbb R}^3$.
The ana\-lysis of ${\mathcal E}_{Z, B{\bf e}_z}^{\rm MTF}$ shows that there exists a minimizer  $\rho_{B{\bf e}_z, N,Z}^{\rm MTF}$ for given $B,N,Z$. Furthermore, such a minimizer `lives on the length scale $\ell$' with 
\begin{align}
\label{eq:ell}
\ell := Z^{-1/3}(1+B/Z^{4/3})^{-2/5} \quad \text{ for }\quad 0 \leq B \ll Z^3,
\end{align}
in the sense that the scaled density,
$$
Z^{-2} (1 + B/Z^{4/3})^{-6/5} \rho^{\rm MTF}_{B{\bf e}_z,N,Z}(\ell x),
$$
has a weak limit as $Z \rightarrow \infty$, $B/Z^3 \rightarrow 0$ with $N/Z$ fixed and $B/Z^{4/3}$ tending to a limit $\beta_{\infty} \in [0, +\infty]$.
This length scale is 
in agreement with the heuristic calculations in Section~\ref{Intro} (denotes $R$ there).
The weak limit thus obtained is the minimizer of a natural limiting functional, but we will not use that fact here.

\begin{thm}
\label{thm:summarize_improved}~\\
Let ${\bf a}_{sc} \in C_0^{\infty}({\mathbb R}^3, {\mathbb R}^3)$
and define 
$$
{\bf a}(x) := \ell {\bf a}_{sc}(x/\ell), \quad \text{ with }\quad\ell := Z^{-1/3}(1+B/Z^{4/3})^{-2/5}.
$$
Let $\lambda>0$, $\beta_{\infty} \in [0, +\infty]$, let $(N,Z,B) = (N_n, Z_n, B_n)$ be a sequence such that
\begin{align*}
\lambda &= N/Z , & 
Z&\rightarrow \infty, &  
B Z^{-4/3}&\rightarrow \beta_{\infty},&
B Z^{-3}  &\rightarrow 0,
\end{align*}
and let $\psi=\psi_{N,Z,B}$ be a sequence of ground states of $H(N,Z,B)$ satisfying 
\begin{align}
\label{eq:Symmetry}
T\psi =\psi,\quad \text{ or } \quad T\psi =-\psi,
\end{align}
then
\begin{align}
\label{eq:result_current}
\frac{1}{{\mathcal E}(Z,B)} \left| \langle \psi, J({\bf a}) \psi \rangle -
\frac{d}{dt} {\Big (} E^{{\rm MTF}}(N,Z,B{\bf e}_z+t B \curl {\bf
a}) \Big)_{t=0}\right| \rightarrow 0 .
\end{align}
\end{thm}

Informally stated, when $N\approx Z$ and $B \ll Z^3$, the quantum mechanical current, 
$$
\langle \psi, J({\bf a}) \psi \rangle,
$$ 
coincides to leading order with the current in MTF-theory, 
$$
\frac{d}{dt} {\Big (} E^{{\rm MTF}}(N,Z,B{\bf e}_z+t B \curl {\bf
a}) \Big)_{t=0}.
$$ 

\begin{remark}~\\
The MTF-current 
$$
\frac{d}{dt} {\Big (} E^{{\rm MTF}}(N,Z,B{\bf e}_z+t B \curl {\bf
a}) \Big)_{t=0},
$$ 
is generally of the same order of magnitude as ${\mathcal E}(Z,B)$. Therefore the two terms on the left in \eqref{eq:result_current} are generally of higher order than their difference.
\end{remark}

Partial results on the current of large atoms have been obtained in previous work.
It was proved in \cite{fournais4} that MTF-theory correctly gives the current in the regime of non-dominant fields $B Z^{-4/3} \leq C$. 
That result was later extended in \cite{fournais8} to allow for magnetic fields strong enough to confine to the lowest Landau band---the precise restriction imposed on the magnetic field strength being for technical reasons $B\ll Z^{\frac{98}{51}}$.
It is the objective of the present paper to 
extend the validity of 
this last result to the entire MTF-region, $B \ll Z^3$,
thus proving that, when MTF-theory correctly predicts the ground state energy, it also correctly predicts the ground state current. The main improvements of the present paper over \cite{fournais8} is a much more precise estimate on confinement to the lowest Landau band (given in Theorem~\ref{thm:loc_improved} below), and the combination of that estimate---in the calculation of the current---with even better confinement estimates to slightly larger subspaces, Theorem~\ref{thm:BiggerConf}.

\subsection{Confinement}~\\
The word {\it confinement} in the title of this paper refers to confinement to the
lowest Landau band. We proceed by properly defining this notion, which has already been informally invoked in the previous discussion.

The kinetic energy operator in the coordinates
perpendicular to the magnetic field, $\hat{K}$, is defined by
\begin{align}
\label{eq:Khat}
\hat{K}
= p_{{\bf A},1}^2+p_{{\bf A},2}^2 +\sigma\cdot {\bf B} = p_{{\bf A},1}^2+p_{{\bf A},2}^2 +B \sigma_3.
\end{align}
By an explicit calculation one sees that $\hat{K}$ is unitarily
equivalent to a harmonic oscillator and
that the spectrum of $\hat{K}$ is $2B ({\mathbb N}\cup\{0\})$.
The {\it lowest Landau band} (for one electron) is defined as the
kernel of the operator $\hat{K}$ (acting on $L^2({\mathbb R}^3,
{\mathbb C}^2)$).
The projection $\Pi_0$ on the lowest Landau band for one
electron has the integral kernel (see \cite{FGPY})
\begin{equation}
   \label{eq:Pi0}
   \Pi_0(x,y) = \frac{B}{2\pi} e^{iB (x_1 y_2-x_2 y_1)/2}
e^{-B(x_{\perp}-y_{\perp})^2/4}
\delta(x_3-y_3) P^{\downarrow},
\end{equation}
where $x_{\perp}=(x_1,x_2)$, and where
$
P^{\downarrow} = \left(\begin{array}{cc} 0&0\\0&1 \end{array}\right),
$
is the projection to the spin-down subspace.
It is clear from the explicit expression for $\Pi_0(x,y)$ (or even from the definition of $\hat{K}$) that the {\it magnetic length scale} is of order $B^{-1/2}$. This will be very important in the later calculations.

We also define $\Pi_{>} = 1- \Pi_0$.
The projections $\Pi_0$ and $\Pi_{>}$ depend on the parameter
$B$ (the strength of  the magnetic field). We will sometimes (after scaling)
need to use $\Pi_0, \Pi_{>}$ for other values of the parameter than the $B$ appearing in the Hamiltonian. In
that case we include explicitly the dependence of $\Pi_0$ on $B$ in the notation by writing 
$\Pi_0(B)$.

We define $\Pi_0^N$ as the projection in
$\wedge_{j=1}^N L^2({\mathbb R}^3, {\mathbb C}^2)$ to the space where
all electrons are in the lowest Landau band, i.e.
\begin{align}
\label{eq:PiN}
\Pi_0^N := \otimes_{j=1}^N \Pi_0^{(j)}.
\end{align}
Now we can define the ground
state energy for electrons in the lowest Landau band
$E_{{\rm conf}}^{\rm{Q}}(N,Z,B)$ as

\begin{align*}
   E_{{\rm conf}}^{\rm{Q}}(N,Z,B) = \inf \Spec \Pi_0^N H(N,Z,B)
\Pi_0^N
   = \!\!\!\!
\inf_{\psi \in \Ran \,\Pi_0^N \setminus \{ 0 \} }
   \frac{\langle \psi, H(N,Z,B) \psi \rangle}{\|\psi\|^2}.
\end{align*}

Our notion of confinement is that the ground state energy for electrons
restricted to the lowest Landau band, $E_{{\rm conf}}^{\rm{Q}}$, and the unrestricted ground state energy, $E^{{\rm Q}}$, are equal to leading order. 
Here `to leading order' holds in an asymptotic regime (in $B, Z$) to be specified in Theorem~\ref{thm:loc_improved} below. 
Let us introduce the parameter $\beta$,
\begin{align}
\beta := B/Z^{4/3}.
\end{align}
In \cite{LSY1} it was proved that confinement holds under the
condition $\beta \rightarrow \infty$. (It also follows from that paper that if the condition $\beta\rightarrow \infty$ is not satisfied, then confinement cannot hold.) The result from \cite{LSY1},
though sufficient for their purposes, does not include a precise
estimate on the remainder term. A first precision of that remainder was
contained in \cite{fournais8}. Here we sharpen that estimate.

\begin{thm}[Confinement to lowest Landau band]
\label{thm:loc_improved}~\\
Let $\lambda> 0$ be given. There exists
$C_1, C_2 > 0$ such that if
$$
N/Z = \lambda,\quad
Z \geq C_2,\quad
\text{ and } \quad\beta = B Z^{-4/3} \geq C_2,
$$
then
\begin{eqnarray}
\label{eq:conf_precise}
   E_{{\rm conf}}^{\rm{Q}}(N,Z,B) \geq E^{\rm{Q}}(N,Z,B) \geq E_{{\rm
conf}}^{\rm{Q}}(N,Z,B) (1-C_1 {\mathcal R}_1),
\end{eqnarray}
where
\begin{align}
\label{eq:R1}
{\mathcal R}_1 =  \begin{cases}\min\big(\beta^{-9/10} + \beta^{-9/35} Z^{-2/7}, \beta^{-3/5}\big), & B \leq 2Z^3,\\ \min\big( B^{-1/3}, \frac{Z}{\sqrt{B}}\big), & B \geq 2Z^3.
\end{cases}
\end{align}
\end{thm}

The estimate \eqref{eq:conf_precise} on the total energy implies a strong estimate on the
perpendicular kinetic energy in the ground state.
Given the kinetic energy operator in the coordinates
perpendicular to the magnetic field, $\hat{K}$, defined in \eqref{eq:Khat} above, we
define the corresponding total perpendicular kinetic energy operator as
$$
\hat{K}^N = \sum_{j=1}^N \hat{K}^{(j)}.
$$
Under the same assumptions as in Theorem~\ref{thm:loc_improved}, we can now estimate the expectation of $\hat{K}^N$ in the ground state by the error term ${\mathcal R}_1$.

\begin{cor}
\label{cor:kinetic}~\\
Let the assumptions and notations be as in Theorem~\ref{thm:loc_improved}.
Suppose that $\psi = \psi_{N,Z,B}$ is a ground state for $H(N,Z,B)$.
Then
the perpendicular kinetic energy satisfies the estimate
\begin{eqnarray}
\label{eq:KhatR1}
   0 \leq \langle \psi \,,\, \hat{K}^N  \psi \rangle \leq C_1 {\mathcal R}_1
{\mathcal E}(Z,B),
\end{eqnarray}
where ${\mathcal R}_1$ denotes the error term from \eqref{eq:R1}.
\end{cor}

\subsection{Organisation of the paper}~\\
The main part of the paper is devoted to proofs of confinement to the lowest Landau band and to some slightly larger spaces. This analysis contains the principal new ideas. Sections~\ref{current}--\ref{KIN} contain a discussion of the current and the reduction of the proof of Theorem~\ref{thm:summarize_improved} to such confinement estimates which is our motivation for the present work. 
We recall the main steps in the calculation of the current in order to make the paper reasonably self-contained.
If one is willing to accept the results of Corollary~\ref{cor:kinetic} and Corollary~\ref{cor:Biggerkinetic} and mainly interested in the current, it is possible to jump directly to Section~\ref{current}. \\
In Section~\ref{basic} we recall estimates needed in the further analysis, mainly of Lieb-Thirring type.
The important Section~\ref{LocPi0} contains the proof of the result on confinement announced in Theorem~\ref{thm:loc_improved}.
In Section~\ref{larger} we use the same type of analysis as in Section~\ref{LocPi0} to prove that much better estimates on localisation can be obtained if one replaces the lowest Landau band by a slightly larger space.

~

\noindent {\bf Thanks.}~\\
The author would like to thank B.~Helffer and J.~P.~Solovej for discussions on this subject and comments on preliminary versions of the article.

\section{Useful inequalities}
\label{basic}

In this section we recall the inequalities of Lieb-Thirring type that
will be used in the proof of Theorem~\ref{thm:loc_improved}. Furthermore, we state some basic estimates on parts of the energy (kinetic or potential) that will be used as {\it a priori} input in calculations. We also give the Lieb-Oxford inequality which will be used in Sections~\ref{Jdens} and~\ref{Jint}.

For a self-adjoint operator $A$ such that $\Spec(A) \cap (-\infty, 0)$ is discrete, we write $\{e_j(A)\}_{j=1}^K$, 
with $K \in {\mathbb N} \cup \{ + \infty\}$, for the non-decreasing sequence of negative eigenvalues (counted with multiplicities). The operators considered in Proposition~\ref{prop:LT} below will all satisfy this assumption.
\pagebreak[3]
\begin{prop}[Lieb-Thirring inequalities]
\label{prop:LT}
$\,$
\begin{enumerate}[\normalfont (i)]
\item
\label{LT-classic}
Let $V \in L^{1+d/2}({\mathbb R}^d)$, then
\begin{equation}
\label{eq:LT}
\sum_j e_j({\bf p}^2 + V) 
\geq - C_d \int [V(x)]_{-}^{1+d/2}\,dx,
\end{equation}
where $C_d$ is some positive constant depending only on the
dimension $d$.
\item
\label{LT-mag_classic}
The inequality \eqref{eq:LT} remains true with the same constant if
we replace ${\bf p}$ by ${\bf p}_{{\bf A}}$.
\item
\label{LT-mag_lowest}
If $V \in L^{3/2}({\mathbb R}^3)$, then
\begin{equation}
\label{eq:LT_mag_0}
\sum_j e_j\big( \Pi_0(H_{{\bf A}} + V)\Pi_0\big)
\geq - \frac{2}{3\pi} B
\int [V(x)]_{-}^{3/2}\,dx.
\end{equation}
\item
\label{LT-mag}
If $V \in L^{3/2}({\mathbb R}^3) \cap L^{5/2}({\mathbb
R}^3)$, then
\begin{equation}
\label{eq:LT_mag}
\sum_j e_j (H_{{\bf A}} + V) \geq - \frac{4}{3\pi} B
\int [V(x)]_{-}^{3/2}\,dx
-
\frac{8 \sqrt{6}}{5\pi}
\int [V(x)]_{-}^{5/2}\,dx.
\end{equation}
\end{enumerate}
\end{prop}

\begin{proof}
The estimate in (\ref{LT-classic}) was first proved in
\cite{Lieb-Thirring}. The extension in (\ref{LT-mag_classic})
(based on the diamagnetic inequality) can
for instance be found in \cite{Simon}. Finally, the truly
magnetic estimates (\ref{LT-mag_lowest}), (\ref{LT-mag}),
these are
\cite[Theorem 2.7]{LSY} and \cite[Theorem 2.1]{LSY},
respectively.
\end{proof}

\begin{remark}~\\
There exist generalisations to non-constant magnetic fields. See \cite{ES,ES1,ES3,ES4} and references therein.
\end{remark}

Next we give, without proof, a result on magnitudes of different parts of the energy. This result follows from \cite{LSY1, LSY}.

\begin{prop}
\label{prop:apriori}~\\
Let $\lambda>0$ and be given. Then there exists a constant $C>0$ such that for all $N,Z,B$ with 
$$
 N/Z = \lambda, \quad Z \geq C,
$$ 
and all normalised ground states $\psi$ of $H(N,Z,B)$ 
we have
$$
\big\langle \psi, \sum_{j=1}^N (p_3^{(j)})^2 \psi \big\rangle\leq C {\mathcal E}(Z,B).
$$
Furthermore, if $\phi \in C_0({\mathbb R}^3)$, and $B \leq 2Z^3$ then, with $\ell = Z^{-1/3}(1 + B/Z^{4/3})^{-2/5}$, we can choose $C$ such that also
$$
\Big|\big\langle \psi, \sum_{j=1}^N Z \ell^{-1} \phi(x^{(j)}/\ell) \psi  \big\rangle \Big| \leq C {\mathcal E}(Z,B).
$$
Finally, this last estimate remains true when restricted to the lowest Landau band
$$
\Big|\big\langle \psi, \sum_{j=1}^N Z \ell^{-1} \Pi_0^{(j)} \phi(x^{(j)}/\ell) \Pi_0^{(j)} \psi  \big\rangle \Big| \leq C {\mathcal E}(Z,B).
$$
\end{prop}

For very large $B$, i.e. $B \gg Z^3$, the Lieb-Thirring inequalities are too expensive. Then we need the following bound \cite[2.5 Theorem]{LSY}.

\begin{thm}
\label{thm:Hydrogen}~\\
There exists a constant $C>0$, such that for all $N,Z,B>0$,
$$
\sum_{j=1}^N e_j\big( \Pi_0 \{ H_{{\bf A}} -\frac{Z}{|x|} \} \Pi_0 \big) \geq
-C NZ^2 \big( [\log(\tfrac{B}{Z^3} + 1)]^2+1\big).
$$
\end{thm}

For $\psi \in L^2({\mathbb R}^{3N})$ denote by $\rho_{\psi} \in L^1({\mathbb R}^3)$ the corresponding density
$$
\rho_{\psi}(x) := \sum_{j=1}^N \int_{{\mathbb R}^{3N}} |\psi(x^{(1)}, \ldots, x^{(N)})|^2 \delta(x-x^{(j)}) \,dx^{(1)} \cdots dx^{(N)}.
$$
With this notation the following correlation inequality holds \cite{Lieb-Oxford}.

\begin{thm}[Lieb-Oxford inequality]\label{thm:LO}~\\
There exists a constant $C_{LO}>0$ satisfying that if $\psi \in L^2({\mathbb R}^{3N})$ is normalised such that $\| \psi \|_{L^2} = 1$, then
$$
\langle \psi, \sum_{j<k} \frac{1}{|x^{(j)} -x^{(k)}|} \psi \rangle \geq D(\rho_{\psi}, \rho_{\psi}) - C_{LO} \int_{{\mathbb R}^3} \rho_{\psi}^{4/3}\,dx,
$$
where
$D(f,g) = \iint_{{\mathbb R}^3 \times {\mathbb R}^3} \overline{f(x)} |x-y|^{-1} g(y) \,dxdy$.
\end{thm}

\section{Proof of Theorem~\ref{thm:loc_improved}}
\label{LocPi0}
The proof will be based on the Lieb-Thirring inequalities
from Section \ref{basic} and on estimates on the commutator
between $\Pi_0$ and the Coulomb potential.

\subsection{Estimates on commutators}~\\
One basic input to the proof of Theorem~\ref{thm:loc_improved}
is the following easy lemma.

\begin{lemma}
\label{lem:com_first}~\\
There exists a constant $c>0$ such that for all $f \in
C({\mathbb R}^3)$ with $\nabla f \in L^{\infty}({\mathbb
R}^3)$, we have
$$
\big\| [ \Pi_0(1), f ] \big\| \leq c \| \nabla f\|_{\infty}.
$$
\end{lemma}

Remember that the notation $\Pi_0(1)$ denotes the projection from \eqref{eq:Pi0} with the parameter $B=1$. Putting back the $B$'s (i.e. performing a scaling) we get informally the relation
\begin{align}
\label{eq:informal}
\big[\Pi_0, \frac{1}{|x|} \big] \approx B^{-1/2} \frac{1}{|x|^2},
\end{align}
suggesting that $\Pi_0$ essentially commutes with the Coulomb potential away from a very small neighbourhood of the origin.
This, as also used in \cite{fournais8}, is our main new idea compared to \cite{LSY1}. In \cite{fournais8} a fixed scale was introduced to bound the commutator by a constant, whereas in the present paper we will rather bound the commutator with a potential, essentially like the right hand side of \eqref{eq:informal} (see Lemma~\ref{lem:com2} for the precise statement). This turns out to give a much improved confinement estimate.

\begin{proof}[Proof of Lemma~\ref{lem:com_first}]~\\
The operator $K = [ \Pi_0(1), f ]$ has integral kernel
$$
K(x,y) = \Pi_0(1)(x,y) \big(f(y) -f(x)\big).
$$ 
We estimate the norm
of $K$ by `Schur's Lemma'
\begin{align}
\label{eq:Schur}
\| K \| \leq \max \big\{ \sup_x \int_{{\mathbb R}^3} |K(x,y)| \,
dy, \sup_y \int_{{\mathbb R}^3} |K(x,y)| \, dx \big\},
\end{align}
using that $|f(y) -f(x)| \leq |x-y| \| \nabla f\|_{\infty}$.
\end{proof}

To state the next lemma we introduce a bit of notation. 
First of all we will denote the positive Coulomb potential by $V$:
\begin{align}
\label{eq:Coulomb}
V(x) := \frac{1}{|x|}.
\end{align}
Let
$f_1, f_2$ be a smooth partition of unity on ${\mathbb R}$:
\begin{align*}
&f_1^2(t) + f_2^2(t) = 1, && f_1(t) =1 \text{ for all }
t\leq 1/2,  \\
& f_1  \text{ is
everywhere non-increasing},&& f_1(t)  = 0 \text{ for all } t\geq 1.
\end{align*}
Define the cut-off Coulomb potentials by
\begin{align*}
v_{<}(x) = \frac{f_1^2(|x|)}{|x|},\quad\quad v_{>}(x) = \frac{f_2^2(|x|)}{|x|},\quad\quad x \in {\mathbb R}^3.
\end{align*}
Define furthermore, with $\langle x \rangle = \sqrt{1+x^2}$,  
\begin{align}
\label{eq:V<}
V_{<}(x) &:= B^{1/2} v_{<}(B^{1/2}x) = \frac{f_1^2(B^{1/2}|x|)}{|x|},\\
\label{eq:V>}
V_{>}(x) &:= B^{1/2} v_{>}(B^{1/2}x) = \frac{f_2^2(B^{1/2}|x|)}{|x|},
\intertext{and }
\label{eq:U>}
U_{>}(x) &= \frac{1}{\langle B^{1/2}x \rangle}.
\end{align}

\begin{prop}
\label{prop:com_applicable}~\\
Let $s \in [0,2]$.
With the constant $c$ from Lemma \ref{lem:com2} we have for
all $\psi \in L^2({\mathbb R}^3)$ and for all $\epsilon > 0$,
\begin{align}
\left| \langle \psi, \Pi_0 [\Pi_0, V_{>} ] \Pi_{>} \psi
\rangle \right|
\leq \tfrac{1}{2} c B^{1/2}\left\{  \epsilon \| U_>^{s}\Pi_0
\psi \|^2 + \epsilon^{-1} \| U_>^{2-s} \Pi_> \psi \|^2
\right\}.
\end{align}
\end{prop}

We will generally apply Proposition~\ref{prop:com_applicable} with the choice $s=5/4$. This parameter value assures that $U_>^{2s} \in L^{3/2}({\mathbb R}^3)$ and $U_>^{4-2s} \in L^{5/2}({\mathbb R}^3)$, which will be needed for the Lieb-Thirring estimates. Recall from Proposition~\ref{prop:LT} (\ref{LT-mag_lowest}) that the Lieb-Thirring inequality in the lowest Landau band involves an $L^{3/2}$-norm of the potential, whereas the standard Lieb-Thirring inequality in three dimensions \eqref{eq:LT} (which will be applied in the higher Landau bands) contains an $L^{5/2}$-norm, thus demanding less decay of the potential at infinity.

Proposition~\ref{prop:com_applicable} is an immediate consequence of the following lemma (application of Lemma \ref{lem:com2} to the case $\phi_1 =
\Pi_0 \psi$, $\phi_2 = \Pi_{>} \psi$).

\begin{lemma}
\label{lem:com2}~\\
Let $\Pi_0$ be the projection on the lowest Landau band as
in \eqref{eq:Pi0} and let $s \in [0,2]$. Then there exists a constant $c$
(independent of $B$) such that for all $\phi_1, \phi_2 \in
L^2({\mathbb R}^3)$,
\begin{align}
\label{eq:commutator_with_function}
\left| \langle \phi_1, [\Pi_0, V_{>} ] \phi_2 \rangle
\right| \leq c B^{1/2} \| U^{s}_{>} \phi_1 \| \, \| U^{2-s}_{>}
\phi_2 \|.
\end{align}
\end{lemma}

\begin{proof}[Proof of Lemma \ref{lem:com2}]~\\
By scaling it suffices to prove the lemma for $B=1$, in
which case \eqref{eq:commutator_with_function} becomes
\begin{align}
\left| \langle \phi_1, [\Pi_0(1), v_{>} ] \phi_2 \rangle
\right| \leq c \big\| \frac{1}{\langle x \rangle^{s}} \phi_1 \big\| \,
\big\|\frac{1}{\langle x \rangle^{2-s}} \phi_2 \big\|.
\end{align}
Therefore, it suffices to prove a bound on
$\langle x \rangle^{s} [\Pi_0(1), v_{>} ] \langle x \rangle^{2-s}$ in operator norm, which we do using \eqref{eq:Schur}. Writing $K(x,y)$ for the integral kernel of the operator in question we find, using \eqref{eq:Pi0},
\begin{align*}
|K(x,y)|\leq
\frac{1}{2\pi} \langle x \rangle^{s}\langle y \rangle^{2-s}
e^{-(x_{\perp}-y_{\perp})^2/4} \delta(x_3-y_3) \,\big|v_>(y)-v_>(x)\big|.
\end{align*}
We will use the simple estimates
\begin{align}
&|\nabla v_>(x) | \leq C \langle x \rangle^{-2},\nonumber \\
&\frac{\langle \xi \rangle}{\langle \xi -\eta \rangle} \leq \sqrt{2} \langle \eta \rangle,
\end{align}
and Taylor' formula.
Thereby we find, with $\eta'=(\eta_{\perp},0) = (y_{\perp},x_3)-(x_{\perp},x_3)$,
\begin{align*}
\int_{{\mathbb R}^3} |K(x,y)| \, dy
&\leq
C\int_{{\mathbb R}^2} \int_0^1\langle x \rangle^{s}\langle x+\eta' \rangle^{2-s}
e^{-\eta_{\perp}^2/4}  \langle x + t\eta' \rangle^{-2}\,dt \,d\eta_{\perp}\\
&\leq
2C \int_{{\mathbb R}^2} \int_0^1 \langle t \eta' \rangle^s \langle (1-t) \eta'\rangle^{2-s}
e^{-\eta_{\perp}^2/4} \,dt \,d\eta_{\perp} <+\infty,
\end{align*}
uniformly in $x$.
The estimate with the roles of $x$ and $y$ inverted is similar.
\end{proof}

Armed with Proposition~\ref{prop:com_applicable} we can now
proceed to the proof of Theorem \ref{thm:loc_improved}. 
Remember that the ground state energy has the order of magnitude
$$
E \approx Z^{7/3} \beta^{2/5} = B^{2/5} Z^{9/5},
$$ 
up to $B = 2Z^3$, and 
$$
E \approx Z^3 \big(\log \frac{B}{Z^3}\big)^2,
$$
for $B \geq 2Z^3$. In order to keep track of this difference, we divide the proof in two.

\subsection{The MTF-regime: $B \leq 2Z^3$}~\\
We first
consider the case without electron-electron repulsion. The
proof in this simpler case contains all the new ideas (compared to \cite{LSY1,fournais8}) needed for
the full atomic case and we consider that the choices of
parameters come out clearer in this case. Afterwards we
explain the extra argument necessary to handle the
two-particle interaction. 

{\bf Case 1. No two-particle terms.}

In this case the Pauli Hamiltonian \eqref{eq:hamiltonian}
is
\begin{align}
   \label{eq:hamiltonian2}
   H(N,Z,B) = &\sum_{j=1}^N \left( H_{{\bf A}}^{(j)}
-\frac{Z}{|x^{(j)}|} \right).
\end{align}
Remember the definitions of $\Pi_0$ from \eqref{eq:Pi0} and that $\Pi_> = 1 - \Pi_0$.
For all subsets $\alpha\subseteq \{1,\ldots,N\}$, we write
$\tilde{\alpha} = \{1,\ldots,N\} \setminus \alpha$ and define
$$
\Pi^{\alpha} = \prod_{j \in \alpha} \Pi_0^{(j)}
\prod_{k \in \tilde{\alpha}} \Pi_{>}^{(k)} .
$$
Then, with the constant $c$ from Lemma \ref{lem:com2} and all
$\epsilon > 0$, we will prove below the inequality (as operators
on
$\otimes_{j=1}^N L^2({\mathbb R}^3, {\mathbb C}^2)$):
\begin{eqnarray}
\label{eq:decomp_new}
H(N,Z,B) \geq \sum_{\alpha} \Pi^{\alpha}\left( \hat{H}^{\alpha} +
\tilde{H}^{\alpha} \right) \Pi^{\alpha},
\end{eqnarray}
where, with the notation for the potentials introduced in \eqref{eq:Coulomb}, \eqref{eq:V<}, \eqref{eq:U>} and \eqref{eq:V>},
\begin{align*}
\hat{H}^{\alpha} &= \sum_{j\in \alpha} \Big(
H_{{\bf A}}^{(j)} -Z V(x^{(j)})
- {\mathcal C} \epsilon Z B^{1/2} U_{>}^{5/2}(x^{(j)})
-\epsilon Z V_{<}(x^{(j)}) \Big).
\end{align*}
and
\begin{align*}
\tilde{H}^{\alpha} =& \sum_{j\notin \alpha}
\Big( H_{{\bf A}}^{(j)} -Z V(x^{(j)})
-{\mathcal C} \epsilon^{-1} Z B^{1/2} U_{>}^{3/2}(x^{(j)})
-\epsilon^{-1} Z V_{<}(x^{(j)}) \Big),
\end{align*}
and ${\mathcal C}\geq 1$ is a constant that will be specified below.

The inequality \eqref{eq:decomp_new} is an equality for the
kinetic energies $H_{{\bf A}}^{(j)}$ since these commute
with the projections $\Pi^{\alpha}$. For the potential terms
the inequality follows from the decomposition
\begin{align*}
- \frac{1}{|x|} & = - \Pi_0 \frac{1}{|x|} \Pi_0
- \Pi_> \frac{1}{|x|} \Pi_> \\
& \quad-
\big\{ \Pi_0 (V_{<}(x) + V_{>}(x)) \Pi_{>}
+ \Pi_{>} (V_{<}(x) + V_{>}(x)) \Pi_0
\big\},
\end{align*}
the inequalities (Cauchy-Schwarz)
\begin{align*}
\Pi_0 V_{<}(x) \Pi_{>} + \Pi_{>} V_{<}(x) \Pi_0 \leq
\epsilon \Pi_0 V_{<}(x) \Pi_0 +
\epsilon^{-1} \Pi_{>} V_{<}(x) \Pi_{>},
\end{align*}
and (c.f. Proposition~\ref{prop:com_applicable})
\begin{align*}
\Pi_0 V_{>} \Pi_{>} + \Pi_{>} V_{>} \Pi_0 & =
\Pi_0 [ \Pi_0, V_{>}] \Pi_{>} + \Pi_{>} [V_{>}, \Pi_0] \Pi_0
\\
& \leq
{\mathcal C}\epsilon B^{1/2}\Pi_0 U^{5/2}_{>} \Pi_0 +
{\mathcal C}\epsilon^{-1} B^{1/2}\Pi_{>} U^{3/2}_{>} \Pi_{>}.
\end{align*}
Thus we have established \eqref{eq:decomp_new}.

We now estimate $\Pi^{\alpha}\hat{H}^{\alpha} \Pi^{\alpha}$
and $\Pi^{\alpha} \tilde{H}^{\alpha} \Pi^{\alpha}$
independently. Notice that if $\psi \in \wedge_{j=1}^N
L^2({\mathbb R}^3, {\mathbb C}^2)$, then $\Pi^{\alpha} \psi$
is separately antisymmetric in the variables in $\alpha$ and
those in $\tilde{\alpha}$.

To estimate $\Pi^{\alpha}\hat{H}^{\alpha} \Pi^{\alpha}$ we
write for $\eta>0$,
\begin{align}
\hat{H}^{\alpha} = (1-2\eta) \sum_{j\in \alpha} \Big( H_{{\bf
A}}^{(j)} - Z  V(x^{(j)}) \Big)
+ \eta \hat{H}^{\alpha}_1
+ \eta \hat{H}^{\alpha}_2,
\end{align}
with
\begin{align}
\label{eq:H-2-hat}
\hat{H}^{\alpha}_1 & = \sum_{j\in \alpha} H_{{\bf A}}^{(j)}
- 2Z V(x^{(j)})
\nonumber \\
\hat{H}^{\alpha}_2 & = \sum_{j\in \alpha} H_{{\bf A}}^{(j)}
- \eta^{-1}
\epsilon Z B^{1/2} W_1(x^{(j)}),
\end{align}
with
\begin{align}
\label{eq:w1}
W_1(x):=w_1(B^{1/2}x),\quad
w_1(x):= \frac{f_1^2(x)}{|x|} + {\mathcal C}\frac{1}{\langle x \rangle^{5/2}}.
\end{align}

The operator $\hat{H}^{\alpha}_1$ is an atomic Schr\"{o}dinger operator (without the terms corresponding to the electronic repulsion) with nuclear charge $2Z$ and $|\alpha|$ electrons. By the easy part of the HVZ-theorem we get a lower energy by adding electrons, so we may, for a lower bound, assume that $|\alpha| =N$. But for $|\alpha|=N$ we have
$$
\inf \Spec \hat{H}^{\alpha}_1 = E^Q(N,2Z,B).
$$

Notice that we here implicitly used the remark above on the symmetry properties of $\Pi^{\alpha}$ since we considered $\hat{H}^{\alpha}_1$ as an operator on the antisymmetric space $\wedge_{j \in \alpha} L^2({\mathbb R}^3, {\mathbb C}^2)$. We will make similar estimates without repeating this remark.

We therefore get, since ${\mathcal E}(2Z,B)/{\mathcal E}(Z,B)$ is bounded, 
\begin{align}
\inf \Spec \hat{H}^{\alpha}_1 \geq -C {\mathcal E}(Z,B),
\end{align}
for all $\alpha$.

For $\hat{H}^{\alpha}_2$ we can estimate, using Proposition~\ref{prop:LT} (\ref{LT-mag_lowest}) (i.e. the
Lieb-Thirring inequality in the lowest Landau band) and observing that $w_1 \in L^{3/2}({\mathbb R}^3)$:
\begin{align}
\label{eq:lowerh2hat}
\Pi^{\alpha} \hat{H}^{\alpha}_2 \Pi^{\alpha} & \geq -c  B  \int
[ \eta^{-1} \epsilon Z B^{1/2} W_1(x)]^{3/2} \,dx\nonumber\\
& \geq -c B^{1/4} \eta^{-3/2}
\epsilon^{3/2}  Z^{3/2}  \int_{{\mathbb R}^3} w_1^{3/2}(x)\,dx
= -c' \eta^{-3/2}
\epsilon^{3/2} Z^{3/2}
\beta^{1/4} Z^{1/3} \nonumber \\
&=
-C  \eta^{-3/2} \epsilon^{3/2} \beta^{-3/20}
Z^{-1/2}  {\mathcal E}(Z,B).
\end{align}
So our total estimate on $\Pi^{\alpha} \hat{H}^{\alpha}
\Pi^{\alpha}$ becomes
\begin{multline}
\label{eq:hatslut}
\Pi^{\alpha} \hat{H}^{\alpha} \Pi^{\alpha}
\geq
\Pi^{\alpha} \Big( (1-2\eta)E^Q_{{\rm conf}}(N,Z,B) \\- C\eta
\big(1 + \eta^{-3/2} \epsilon^{3/2}  \beta^{-3/20} Z^{-1/2} \big)
{\mathcal E}(Z,B) \Big).
\end{multline}

To estimate $\tilde{H}^{\alpha}$ we start by observing
that $\Pi_{>} H_{{\bf A}} \Pi_{>} \geq \frac{1}{2} \Pi_{>}
({\bf p}_{{\bf  A}}^2 + B ) \Pi_{>}$. Therefore, we have to
estimate
\begin{align}
\label{eq:EkstraBs}
\Pi^{\alpha}
\tilde{H}^{\alpha} \Pi^{\alpha}
\geq
\frac{1}{2} \Pi^{\alpha} \left( \tilde{H}_1^{\alpha} +
\tilde{H}_2^{\alpha} + \frac{B
|\tilde{\alpha}|}{2}\right)\Pi^{\alpha},
\end{align}
where
\begin{align}
\tilde{H}^{\alpha}_1 & = \sum_{j \in \tilde{\alpha}}
\frac{1}{2} ({\bf p}_{{\bf A}}^{(j)})^2  +\frac{B}{4} - Z
V(x^{(j)}), \nonumber
\\
\tilde{H}^{\alpha}_2 & = \sum_{j \in \tilde{\alpha}}
\frac{1}{2} ({\bf p}_{{\bf A}}^{(j)})^2  +\frac{B}{4}- 
\epsilon^{-1} Z B^{-1/2}  W_2(x^{(j)}),
\end{align}
with
\begin{align}
W_2(x):= w_2(B^{1/2}x), \quad
w_2(x) :=  \frac{f_1^2(x)}{|x|} + {\mathcal C}\frac{1}{\langle x \rangle^{3/2}}.
\end{align}
Applying the magnetic Lieb-Thirring inequality, Proposition~\ref{prop:LT}~(\ref{LT-mag_classic}), we get
\begin{align}
\label{eq:1tilde}
\tilde{H}^{\alpha}_1 \geq -C \int_{{\mathbb R}^3} \Big[
B-\frac{Z}{|x|} \Big]_{-}^{5/2} \,dx &=  -CB^{-1/2} Z^3 \int_0^1 (1-\rho^{-1})^{5/2}\rho^2\,d\rho \nonumber\\
&=-C' \beta^{-9/10} {\mathcal E}(Z,B),
\end{align}
and, using $w_2 \in L^{5/2}({\mathbb R}^3)$,
\begin{align}
\label{eq:tildeUnderneath}
\tilde{H}^{\alpha}_2 & \geq
  -C \int_{ {\mathbb R}^3 }  \big[ B- \epsilon^{-1} Z
B^{1/2}W_2(x)  \big]_{-}^{5/2} \,dx \nonumber \\
&\geq -C \int_{ {\mathbb R}^3 }  \big[ \epsilon^{-1} Z
B^{1/2}W_2(x)  \big]^{5/2} \,dx \nonumber \\
&= -C' \epsilon^{-5/2} Z^{5/2} B^{-1/4} = -C'
\epsilon^{-5/2}  Z^{13/6} \beta^{-1/4} \nonumber \\
  & = -C' \epsilon^{-5/2} Z^{-1/6} \beta^{-13/20} {\mathcal E}(Z,B).
\end{align}
Using \eqref{eq:EkstraBs}, \eqref{eq:1tilde} and \eqref{eq:tildeUnderneath}, we get
\begin{align}
\label{eq:tildeslut}
\Pi^{\alpha}
\tilde{H}^{\alpha} \Pi^{\alpha}
\geq
\Pi^{\alpha} \left( -c\big[\beta^{-9/10} + \epsilon^{-5/2}
Z^{-1/6} \beta^{-13/20}\big] {\mathcal E}(Z,B) + \frac{|\tilde{\alpha}|}{4} B \right).
\end{align}

{\bf First choice of parameters}\\
For $B \leq Z^{13/6}$ we cannot profit from the {\it positive} term $\frac{|\tilde{\alpha}|}{4} B$ in \eqref{eq:tildeslut}.
Combining \eqref{eq:decomp_new} with \eqref{eq:hatslut} and \eqref{eq:tildeslut} we find
\begin{multline}
\label{eq:ErrBsmall}
E^Q(N,Z,B) \geq (1-2\eta)E^Q_{{\rm conf}}(N,Z,B) \\
- C\Big[\eta
\big(1 + \eta^{-3/2} \epsilon^{3/2}  \beta^{-3/20} Z^{-1/2} \big)\\
+\big(\beta^{-9/10} + \epsilon^{-5/2}
Z^{-1/6} \beta^{-13/20}\big)
\Big]{\mathcal E}(Z,B) .
\end{multline}

To get an optimal choice of parameters,
$\eta, \epsilon$, in \eqref{eq:ErrBsmall} we set the leading
error estimates to be equal:
\begin{align}
\eta^{-3/2} \epsilon^{3/2}  \beta^{-3/20} Z^{-1/2}
&= 1 ,&
\epsilon^{-5/2} Z^{-1/6} \beta^{-13/20} &= \eta.
\end{align}
This implies the choice
\begin{align}
\eta &= \beta^{-9/35} Z^{-2/7} ,  &
\epsilon & = \beta^{-11/70} Z^{1/21}.
\end{align}
With this choice we get the estimate
$$
E^Q \geq E_{{\rm conf}} + R,
$$
with
$$
R = O(\beta^{-9/10}) + O(\beta^{-9/35} Z^{-2/7} ).
$$
This is the first error term in Theorem~\ref{thm:loc_improved} for $B \leq 2Z^3$ and is the better of the two for $B \leq Z^{13/6}$.

{\bf Second choice of parameters} \\
For $B \gtrsim Z^{13/6}$ the positive term in \eqref{eq:tildeslut} dominates the negative terms in that equation---even for $|\tilde{\alpha}|=1$.
Therefore we can make a somewhat more natural choice of parameters, leading to the relative error term in $O(\beta^{-3/5})$ and thus better for large $B$, as follows.

We choose
\begin{align}
\label{eq:epsilon}
\epsilon = M Z B^{-1/2} = M Z^{1/3} \beta^{-1/2},\quad\quad\quad\quad \eta=\beta^{-3/5},
\end{align}
where $M$ is a (large) constant to be chosen below. We start
by analysing $\tilde{H}^{\alpha}$ a
bit differently from above.

The (bottom of the spectrum of the) operator $\tilde{H}^{\alpha}_1$ is estimated as before by
$O(\beta^{-9/10} {\mathcal E}(Z,B))$.
For $\tilde{H}^{\alpha}_2$, we apply the value of $\epsilon$ from \eqref{eq:epsilon} and get
\begin{align}
\tilde{H}^{\alpha}_2 & = \sum_j  \frac{1}{2} ({\bf p}_{{\bf A}}^{(j)})^2
+\frac{B}{4}- M^{-1} W_2(x^{(j)}).
\end{align}
By scaling, $\frac{1}{B} \tilde{H}^{\alpha}_2$ is unitarily equivalent to the $(Z,B)$
independent operator
\begin{align}
\tilde{h}^{\alpha}_2 & = \sum_j  \Big\{\frac{1}{2} ({\bf q}^{(j)})^2 +\frac{1}{4}-
M^{-1} \Big(\frac{f_1^2(x^{(j)})}{|x^{(j)}|}+{\mathcal C} \frac{1}{\langle x^{(j)} \rangle^{3/2}}\Big)\Big\},
\end{align}
where ${\bf q}=(-i\nabla + B^{-1}{\bf A})$ (notice that $\nabla \times (B^{-1} {\bf A}) =(0,0,1)$).
Upon application of the Lieb-Thirring inequality, Proposition~\ref{prop:LT} (\ref{LT-mag_classic}), we see
that
$$
\tilde{h}^{\alpha}_2  \geq -C(M),
$$
where $C(M) \rightarrow 0$ as $M \rightarrow +\infty$. Therefore, remembering the term $\frac{B|\alpha|}{2}$ in \eqref{eq:EkstraBs}, for
some sufficiently large (but fixed) $M$ we find the estimate, for $|\tilde{\alpha}| \neq 0$,
\begin{align}
\label{eq:tildelower}
\Pi^{\alpha} \tilde{H}^{\alpha} \Pi^{\alpha}
\geq
\frac{1}{2} \Pi^{\alpha} \left( -C \beta^{-9/10} {\mathcal E}(Z,B) + \frac{B}{2} (
|\tilde{\alpha}| - \tfrac{1}{10})\right)\Pi^{\alpha}.
\end{align}

When considering $\hat{H}^{\alpha}$ remember the parameter choices from \eqref{eq:epsilon}.  Using the estimate in \eqref{eq:lowerh2hat}
we find
\begin{align}
\Pi^{\alpha} \hat{H}_2^{\alpha}  \Pi^{\alpha} \geq -C_M   {\mathcal E}(Z,B) \Pi^{\alpha}.
\end{align}
Therefore, $\hat{H}^{\alpha}$ satisfies
\begin{align}
\label{eq:hatlower}
\Pi^{\alpha} \hat{H}^{\alpha} \Pi^{\alpha}
\geq
\Pi^{\alpha} \left( (1-3 \beta^{-3/5})E_{{\rm conf}}^Q  -C_M   \beta^{-3/5} {\mathcal E}(Z,B)\right)\Pi^{\alpha}.
\end{align}
So combining \eqref{eq:tildelower} with \eqref{eq:hatlower} (omitting the positive term proportional to $B$), we find,
\begin{align}
\label{eq:totallower}
\Pi^{\alpha} \left( \hat{H}^{\alpha} +  \tilde{H}^{\alpha} \right)
\Pi^{\alpha} 
\geq
\Pi^{\alpha} \left( E_{{\rm conf}}^Q  + 3 \beta^{-3/5}|E_{{\rm conf}}^Q|
   -C_M   \beta^{-3/5}{\mathcal E}(Z,B)
\right)\Pi^{\alpha}. 
\end{align}
This finishes the proof of Theorem \ref{thm:loc_improved} in
the case without two-particle potentials and for parameters $B\leq 2Z^3$.

\begin{remark}~\\
If we include the positive term from \eqref{eq:tildelower}, and use that $|E_{{\rm conf}}^Q|/|E^Q|\rightarrow 1$, we get the following more precise version of \eqref{eq:totallower}
\begin{align}
\label{eq:feschbach2}
\sum_{\alpha} \Pi^{\alpha} \left( \hat{H}^{\alpha} +
\tilde{H}^{\alpha} \right) \Pi^{\alpha} &\geq
\left( E_{{\rm conf}}^Q  
   -C  \beta^{-3/5} {\mathcal E}(Z,B) \right) \Pi_0^N \nonumber\\
&\quad+
\sum_{|\alpha| < N } \big(E_{{\rm conf}}^Q - C \beta^{-3/5}{\mathcal E}(Z,B) + \frac{B}{10} |\tilde{\alpha}|\big) \Pi^{\alpha}.
\end{align}
For $B \gg Z^{13/6}$ we have $B \gg \beta^{-3/5}{\mathcal E}(Z,B)$, and therefore \eqref{eq:feschbach2} implies the existence of a constant $c>0$ such that
\begin{align}
\label{eq:feschbach3}
(1- \Pi_0^N) H (1- \Pi_0^N) \geq \left( E^{{\rm Q}}_{{\rm conf}} + cB \right)(1- \Pi_0^N).
\end{align}
This is the necessary assumption for application of the Feschbach method (see for instance \cite{BFS1, BFS2}) to the present problem. So for these (rather large, i.e. $B \gg Z^{13/6}$) fields that approach might work. As can easily be seen from the estimates below, \eqref{eq:feschbach3} remains true, under the condition $B \gg Z^{13/6}$, when the electron-electron terms are included in the Hamiltonian $H$.
\end{remark}

{\bf Case 2. Full atomic problem.}

We now return to the atomic operator given in
\eqref{eq:hamiltonian}. The analysis starts
similarly to the above, only we get more terms. In
particular, the inequality \eqref{eq:decomp_new} becomes
\begin{eqnarray}
\label{eq:decomp_new2}
H \geq \sum_{\alpha} \Pi^{\alpha}\left( \hat{{\mathfrak H}}^{\alpha} +
\tilde{{\mathfrak H}}^{\alpha} \right) \Pi^{\alpha},
\end{eqnarray}
where, with $r_{j,k}:=x^{(j)}-x^{(k)}$,
\begin{align*}
\hat{{\mathfrak H}}^{\alpha} &= \sum_{j\in \alpha} \Big(
H_{{\bf A}}^{(j)} -Z V(x^{(j)})
-\epsilon Z V_{<}(x^{(j)})
- {\mathcal C} \epsilon Z B^{1/2} U_{>}^{5/2}(x^{(j)}) \\
&\quad+ \sum_{j,k \in \alpha\, : \,j<k} \Big(
V(r_{j,k})-3\epsilon
V_{<}(r_{j,k}) -
3{\mathcal C} \epsilon B^{1/2} U_{>}^{5/2}(r_{j,k}) \Big) \\
&\quad+ \sum_{j \in \alpha, k\in \tilde{\alpha}} \Big(
V(r_{j,k})-
\tfrac{3}{2}\epsilon
V_{<}(r_{j,k}) 
- \tfrac{3}{2}{\mathcal C}
\epsilon
B^{1/2}  U_{>}^{5/2}(r_{j,k})\Big),
\end{align*}
and
\begin{align*}
\tilde{{\mathfrak H}}^{\alpha} =& \sum_{j\in \tilde{\alpha}}
\Big( H_{{\bf A}}^{(j)} -Z V(x^{(j)})
   -\epsilon^{-1} Z V_{<}(x^{(j)}) -{\mathcal C}
\epsilon^{-1} Z B^{1/2}  U_{>}^{3/2}(x^{(j)}) \Big) \\
&\quad+ \sum_{j,k \in \tilde{\alpha} \,:\,j<k} \Big(
V(r_{j,k})
-3\epsilon^{-1} V_{<}(r_{j,k}) 
- {\mathcal C} \epsilon^{-1} B^{1/2} U_{>}^{3/2}(r_{j,k}) \Big)
\\
&\quad +\sum_{j \in \alpha, k\in \tilde{\alpha}} \Big(
V(r_{j,k})-
\tfrac{3}{2}\epsilon^{-1}
V_{<}(r_{j,k}) 
- \tfrac{3}{2} {\mathcal C}
\epsilon^{-1}
B^{1/2}  U_{>}^{3/2}(r_{j,k})\Big).
\end{align*}
The proof of \eqref{eq:decomp_new2} is similar to the proof
of \eqref{eq:decomp_new} and will be omitted.

The idea behind the treatment of the  two-particle
terms is as follows. All estimates are done using
Lieb-Thirring inequalities (this was also the case before).
The two-particle terms---typically $r_{j,k}^{-1}$---each come with a coefficient $1$, but there are $N
\approx Z$ of them, so therefore, in total, they will
contribute with a term of the same order of magnitude as the
corresponding one-particle term---typically $Z r_j^{-1}$---coming from the interaction
with the nucleus.

Let us start by estimating $\tilde{{\mathfrak H}}^{\alpha}$. As before,
we can get a lower bound by replacing $H_{{\bf A}}^{(j)}$
(for $j \in \tilde{\alpha}$) by $\frac{1}{2} (({\bf p}_{{\bf
A}}^{(j)})^2 + B )$. Therefore, we find
\begin{align}
\label{eq:before_LevyLeblond}
   &\Pi^{\alpha} \tilde{{\mathfrak H}}^{\alpha}\Pi^{\alpha}
   \geq
   \frac{1}{2}\Pi^{\alpha} \left(\tilde{{\mathfrak H}}_1^{\alpha} +
\tilde{{\mathfrak H}}_2^{\alpha} + \frac{|\tilde{\alpha}| B}{2}\right)\Pi^{\alpha},
\end{align}
where
\begin{align}
\label{eq:Frak1}
\tilde{{\mathfrak H}}_1^{\alpha}  = & \sum_{j\in \tilde{\alpha}} \Big(\frac{({\bf
p}_{{\bf A}}^{(j)})^2+B}{4} - \frac{Z}{|x^{(j)}|}\Big) + \sum_{j,k \in \tilde{\alpha} \,:\, j<k}
\frac{1}{|r_{j,k}|} +
\sum_{j \in \alpha ,k \in \tilde{\alpha}}
\frac{1}{|r_{j,k}|},
\end{align}
and
\begin{align}
\label{eq:Frak2}
\tilde{{\mathfrak H}}_2^{\alpha}  = & \sum_{j\in \tilde{\alpha}} 
\Big(\frac{({\bf p}_{{\bf A}}^{(j)})^2+B}{4} - \epsilon^{-1} Z B^{1/2}
W_2(x^{(j)})\Big)  \nonumber\\
&-3\epsilon^{-1} B^{1/2} \sum_{j,k \in\tilde{\alpha} \,:\, j<k}
W_2(r_{j,k})
-\tfrac{3}{2}\epsilon^{-1}B^{1/2}
\sum_{j \in \alpha ,k \in \tilde{\alpha}}
W_2(r_{j,k}) .
\end{align}

In the first operator, $\tilde{{\mathfrak H}}_1^{\alpha}$ the
two-particle terms come in with a positive sign and can
therefore be neglected for a lower bound. The other 
operator $\tilde{{\mathfrak H}}_2^{\alpha}$ will
cause us a bit more trouble.

Let $n_{\alpha}=|\alpha|$ and $\tilde{n}_{\alpha}=|\tilde{\alpha}| = N-
n_{\alpha}$. For simplicity of notation, we will renumber
the electrons so that $\alpha = \{1,\ldots,n_{\alpha}\}$,
$\tilde{\alpha} = \{n_{\alpha}+1, \ldots,N\}$.
If $n_{\alpha}=N$ then $\tilde{{\mathfrak H}}_2^{\alpha} = 0$.
If not, we can rewrite $\tilde{{\mathfrak H}}_2^{\alpha}$ as follows (using an idea of L\'{e}vy-Leblond \cite{LevyLeblond}) for $\tilde{n}_{\alpha} \geq 2$ (for $\tilde{n}_{\alpha} =1$ no reformulation of \eqref{eq:Frak2} is necessary),
\begin{align}
\label{eq:levyLe}
\tilde{{\mathfrak H}}_2^{\alpha} &= \frac{1}{\tilde{n}_{\alpha}-1}
\sum_{k=n_{\alpha}+1}^N \Big\{
\sum_{j=n_{\alpha}+1, j\neq k}^N \Big( (({\bf
p}_{{\bf A}}^{(j)})^2+B)/4 - \epsilon^{-1} Z B^{1/2}
W_2(x^{(j)}) \nonumber \\
&\quad\quad\quad-\tfrac{3}{2}\epsilon^{-1} B^{1/2}(\tilde{n}_{\alpha}-1)
W_2(r_{j,k})  -\tfrac{3}{2}\epsilon^{-1}B^{1/2}
\sum_{l \in \alpha}
W_2(r_{j,l}) \Big)
\Big\}
\end{align}
We can estimate $\Pi^{\alpha} \{\cdot \} \Pi^{\alpha}$
(where $\{\cdot \}$ denotes the operator inside $\{ \, \}$ in
\eqref{eq:levyLe}) using Proposition~\ref{prop:LT} (\ref{LT-mag_classic}), by
\begin{align}
\Pi^{\alpha} \{\cdot \} \Pi^{\alpha} &\geq
\inf
-C\int_{{\mathbb R}^3}
\Big[ B - 4\epsilon^{-1} Z
B^{1/2}
W_2(x) -6\epsilon^{-1} (\tilde{n}_{\alpha}-1)
B^{1/2}
W_2(x-w) \nonumber \\
& \quad \quad \quad \quad \quad\quad \quad \quad - 6 \epsilon^{-1}
\sum_{l \in \alpha}
B^{1/2}
W_2(x-z_l) \Big]_{-}^{5/2}\,dx \nonumber \\
& \geq -c \epsilon^{-5/2} Z^{-1/6} \beta^{-13/20}
{\mathcal E}(Z,B),
\end{align}
where the $\inf$ denotes $\inf_{\{{\bf z}=(z_1,\ldots,z_{n_{\alpha}) }
\in {\mathbb R}^{3n_{\alpha}}\}}
\inf_{\{w\in {\mathbb R}^3\}}$. Thus,
\begin{align}
\frac{1}{\tilde{n}_{\alpha}-1}
\sum_{k=n_{\alpha}+1}^N \Pi^{\alpha} \{\cdot \} \Pi^{\alpha}
\geq -c \epsilon^{-5/2} Z^{-1/6} \beta^{-13/20}
{\mathcal E}(Z,B).
\end{align}
So we see by comparison with \eqref{eq:tildeUnderneath} that
the estimate is unchanged by the inclusion of the
two-particle terms into the operator.

The same strategy applies to all the other terms: One writes
the operator in the manner illustrated by \eqref{eq:levyLe}
and applies the appropriate Lieb-Thirring inequality. We
omit the details. The error estimates are invariably of the
same type as those given in the case without two-particle
potentials. Therefore, we can use the same choices of
parameters $\epsilon, \eta$ as before and get the same final
estimate.

This finishes the proof of Theorem \ref{thm:loc_improved} in the case $B \leq 2Z^3$.

\subsection{The very strong field regime: $B\geq 2Z^3$}~\\
Notice that in this case we have ${\mathcal E}(Z,B) = Z^3 (\log \frac{B}{Z^3})^2$ and
\begin{align}
\label{eq:B-Z3}
B \geq cZ^3 \Big(\log\frac{B}{Z^3}\Big)^2.
\end{align}
We keep the choice of $\epsilon$ from the second choice of parameters from the MTF-regime, but choose $\eta=B^{-1/3}$:
$$
\epsilon = \frac{Z}{M\sqrt{B}}, \quad \quad \eta=B^{-1/3}.
$$
Arguing as previously, we get \eqref{eq:decomp_new2} and that $\tilde{{\mathfrak H}}^{\alpha}$ is positive (for $M$ sufficiently large independent of $B,Z$). Using \eqref{eq:B-Z3} this becomes:
There exists $C>0$ such that for all $\alpha \neq \{1,\ldots,N\}$,
\begin{align}
\label{eq:TildeFrak}
\Pi^{\alpha} \tilde{{\mathfrak H}}^{\alpha} \Pi^{\alpha}\geq CB \Pi^{\alpha}.
\end{align}
On the $\hat{{\mathfrak H}}^{\alpha}$ we proceed as before, but with the new choice of $\eta$.
The Lieb-Thirring inequality corresponding to the operator $\hat{H}_2^{\alpha}$ from \eqref{eq:H-2-hat} in the case without two-particle terms becomes
\begin{align}
\Pi^{\alpha}\hat{H}_2^{\alpha} \Pi^{\alpha}\geq -c B \int_{{\mathbb R}^3} \big[ B^{1/3} Z^2 W_1(x)\big]^{3/2}\,dx
\geq -c' Z^3.
\end{align}
As before, we get the same order of magnitude when we include the two-body terms. 
The other parts of $\hat{{\mathfrak H}}^{\alpha}$ are of lower order, and
we therefore get a relative error of order $\eta$, i.e.
\begin{align}
\Pi^{\alpha} \hat{{\mathfrak H}}^{\alpha}\Pi^{\alpha}\geq E^{Q}_{\rm conf}(1+CB^{-1/3}) \Pi^{\alpha}.
\end{align}
This, combined with \eqref{eq:TildeFrak}, gives the first error bound in Theorem \ref{thm:loc_improved} for $B\geq 2Z^3$, i.e. the relative error $B^{-1/3}$.

The second error bound in Theorem \ref{thm:loc_improved}, $\frac{Z}{\sqrt{B}}$, was actually proved though not stated explicitly in \cite{LSY1}. Here the commutation, i.e. the application of Proposition~\ref{prop:com_applicable}, is not needed, only the Cauchy-Schwarz inequality\footnote{Inclusion of the commutation in the argument would only improve the estimate \eqref{eq:R1} by a logarithmic factor.}. We therefore get, instead of \eqref{eq:decomp_new2} the simpler estimate
\begin{align}
H \geq \sum_{\alpha} \Pi^{\alpha} \big( \widehat{\mathcal H}^{\alpha}+ \widetilde{\mathcal H}^{\alpha} \big) \Pi^{\alpha},
\end{align}
with
\begin{align*}
 \widehat{\mathcal H}^{\alpha} := \sum_{j\in \alpha} H_{{\bf A}}^{(j)} - (1+\epsilon)Z V(x^{(j)})
 + \sum_{j,k\in \alpha, j<k} (1-3\epsilon)V(r_{j,k}),
\end{align*}
and
\begin{align*}
\widetilde{\mathcal H}^{\alpha} &:=
\sum_{j\in \tilde{\alpha}} H_{{\bf A}}^{(j)} - (1+\epsilon^{-1})Z V(x^{(j)})
 + \sum_{j,k\in \tilde{\alpha}, j<k} (1-3\epsilon^{-1})V(r_{j,k}) \\
&\quad\quad
 + \sum_{j\in \alpha, k\in \tilde{\alpha}} (1-\tfrac{3}{2}\epsilon-\tfrac{3}{2}\epsilon^{-1})V(r_{j,k}).
\end{align*}
We choose $\epsilon = \frac{ZM}{\sqrt{B}}$ for some large $M$ as before. After scaling and discarding some positive terms we find
\begin{align}
\label{eq:TildeBig}
\Pi^{\alpha}  \widetilde{\mathcal H}^{\alpha} \Pi^{\alpha}
\geq
B \Pi^{\alpha} &\Big\{
\sum_{j\in \tilde{\alpha}}\big( \frac{1}{2} ({\bf p}_{B^{-1}{\bf A}}^{(j)})^2 + \frac{1}{2} - M^{-1} V(x^{(j)})\big) \nonumber\\
&
- 3M^{-1} \sum_{j,k\in \tilde{\alpha}, j<k} V(r_{j,k})  -3M^{-1}\sum_{j\in \alpha, k\in \tilde{\alpha}} V(r_{j,k})
\Big\}\Pi^{\alpha}.
\end{align}
Using the same arguments as before we find that there exists $M_0$ such that for $M\geq M_0$ the operator in $\{\cdot\}$ in \eqref{eq:TildeBig} is positive as an operator on $\otimes_{j\in \tilde{\alpha}} L^2({\mathbb R}^3;{\mathbb C}^2)$ {\it independently} of the parameters $\{x_k\}_{k\in \alpha}$. Thus, for $M$ sufficiently large,
\begin{align}
\Pi^{\alpha}  \widetilde{\mathcal H}^{\alpha} \Pi^{\alpha} \geq 0.
\end{align}
To estimate $\widehat{\mathcal H}^{\alpha}$ we rewrite it as
\begin{align*}
\widehat{\mathcal H}^{\alpha} &=
(1-3\epsilon)  \sum_{j\in \alpha} \Big\{H_{{\bf A}}^{(j)} - Z V(x^{(j)})
+ \sum_{j,k\in \alpha, j<k} V(r_{j,k}) \Big\}\\
&\quad+3\epsilon \sum_{j\in \alpha} \Big\{H_{{\bf A}}^{(j)} - \frac{4}{3}Z V(x^{(j)})
 \Big\}. 
\end{align*}
Applying Theorem~\ref{thm:Hydrogen} to the last term we find
\begin{align}
\Pi^{\alpha}  \widehat{\mathcal H}^{\alpha} \Pi^{\alpha}
&\geq
(1-3\epsilon) E^{\rm Q}_{{\rm conf}}(|\alpha|,Z,B)-
C \epsilon N Z^2 ([\log(B/Z^3)]^2+1)\\
&\geq (1-3\epsilon) E^{\rm Q}_{{\rm conf}}(N,Z,B)-
C \epsilon N Z^2 ([\log(B/Z^3)]^2+1).
\end{align}
Remembering the choice $\epsilon = \frac{ZM}{\sqrt{B}}$ we find the relative error $\frac{Z}{\sqrt{B}}$.

This finishes the proof of Theorem~\ref{thm:loc_improved} in the case $B\geq 2Z^3$.
\qed

\subsection{Estimate on the kinetic energy}~\\
We finish this section by giving the short proof of Corollary~\ref{cor:kinetic}
\begin{proof}
Define $H_{\frac{1}{2}}(N,Z,B) := H(N,Z,B) - \frac{1}{2}\hat{K}^N$ and let $E^{\rm Q}_{\frac{1}{2}}(N,Z,B)$ be the ground state energy of $H_{\frac{1}{2}}(N,Z,B)$. By the variational principle we have
\begin{align}
\label{eq:EasyConvex}
\frac{1}{2}\langle \psi \,,\, \hat{K}^N  \psi \rangle
&= \langle \psi \, ,\, \{ H(N,Z,B) - H_{\frac{1}{2}}(N,Z,B) \} \psi \rangle \nonumber\\
&\leq E^{\rm Q}(N,Z,B) - E^{\rm Q}_{\frac{1}{2}}(N,Z,B) \nonumber\\
&\leq E^{\rm Q}_{\rm conf}(N,Z,B) - E^{\rm Q}_{\frac{1}{2}}(N,Z,B) .
\end{align}
Furthermore the proof of \eqref{eq:conf_precise} holds with only notational change for $H_{\frac{1}{2}}(N,Z,B)$. Thus
\begin{align}
\label{eq:Ehalv}
E^{\rm Q}_{\frac{1}{2}}(N,Z,B) \geq E^{\rm Q}_{\rm conf}(N,Z,B)(1 - C_1' {\mathcal R}_1).
\end{align}
Combining \eqref{eq:EasyConvex} and \eqref{eq:Ehalv} we get \eqref{eq:KhatR1}.
\end{proof}

\section{An estimate on confinement to a larger space}
\label{larger}
The result on confinement in Theorem~\ref{thm:loc_improved} is rather precise. However, for the application to the calculation of the current it is just slightly too crude. As can be seen from the proof of Theorem~\ref{thm:loc_improved}, the main error in the confinement estimate comes from the region near the singularities of the potential. The region in question has essentially the length scale $B^{-1/2}$, which is very small compared to the other length scales of the atom as discussed in subsection~\ref{Intro}. This suggests that introducing a localisation away from the singular region, one can hope for a more precise estimate.
It is reasonable to make such a localisation in the variable $x_3$ (parallel to the magnetic field), in order for the localisation to commute with the projection on the Landau bands.
That is the rationale behind the next result. It turns out to be convenient---in particular when the two-particle interaction is included---to make the localisation in frequency- instead of position-space.

Let $\Pi_0, \Pi_>$ be as previously defined. 
Let 
\begin{align}
\label{eq:L}
L:= \begin{cases}Z^{-2/5} B^{-1/5}, & B \leq 2Z^3, \\
Z^{-1} [\log \frac{B}{Z^3} ]^{-1}, & B \geq 2Z^3,
\end{cases}
\end{align}
be the parallel length scale and define, for $\delta>0$, projections on low and high frequency  spaces
\begin{align}
p_{{\rm hf}}&:= 1_{\{|p_3| \geq \delta^{-1}L^{-1}\}}, &
p_{{\rm lf}}&:= 1_{\{|p_3| < \delta^{-1}L^{-1}\}} = 1 - p_{{\rm hf}},
\end{align}
where $1_{\Omega}$ denotes the characteristic function of the set $\Omega$.
Clearly, $p_{{\rm hf}}$ commutes with $\Pi_0$ and $\Pi_>$.

Intuitively speaking, the frequencies below $\delta^{-1} L^{-1}$ can only probe length-scales above $\delta L$. We want $\delta L$ to be larger than the magnetic length scale $B^{-1/2}$ but shorter than the length $L$ of the `electronic cylinder'.
Notice that 
$$
\frac{B^{-1/2}}{L} = \begin{cases} \beta^{-3/10}, & B \leq 2Z^3, \\
\frac{Z}{\sqrt{B}} \log \frac{B}{Z^3} , & B \geq 2Z^3.
\end{cases}
$$

Define the orthogonal projections $P_0$ and $P_>$ on $L^2({\mathbb R}^3)$ by
\begin{align}
\label{eq:Pbig}
P_0 &= \Pi_0 + p_{{\rm hf}} \Pi_>, &
P_>&:=p_{{\rm lf}} \Pi_>.
\end{align}
One easily sees that 
$$
[P_0, P_>] = 0, \quad\quad P_0 P_>=0,\quad\quad P_0+P_>=1.
$$
By analogy with the localisation to the lowest Landau band, we also define 
\begin{align}
P_0^{N} := \prod_{j=1}^N P_0^{(j)},\quad\quad P_>^N:= 1 - P_0^{N},
\end{align}
and
\begin{align}
E_{\rm conf,hf}(N,Z,B) := \inf \Spec P_0^{N} H(N,Z,B) P_0^{N}.
\end{align}
It is clear that $p_{\rm hf}$ tends to the identity as $\delta \rightarrow \infty$. Thus, we expect to get an improved confinement estimate, compared to Theorem~\ref{thm:loc_improved}, when $\delta$ is chosen `large'. Our result below shows that, when
$Z^2 \ll B $, we can take $\delta$ slightly smaller than unity and get an estimate with almost a factor of $\frac{Z}{\sqrt{B}}$ of improvement over the ${\mathcal R}_1$ in Theorem~\ref{thm:loc_improved}. 

\begin{thm}
\label{thm:BiggerConf}~\\
Let $\lambda>0$ be given. Then for all $\mu \in (0,1/2)$ there exist $C_1, C_2>0$ such that if
\begin{align}
\label{eq:CondParam}
N/Z = \lambda, \quad Z\geq C_2, \quad 
C_2 Z^2 \leq B,
\end{align}
and (with $L$ defined by \eqref{eq:L})
\begin{align}
\label{eq:deltaLower}
C_2 \frac{B^{-1/2}}{L}  \leq \delta \leq C_2^{-1},
\end{align}
then
\begin{align}
\label{eq:NewConf}
E_{\rm conf,hf}(N,Z,B) \geq 
E^Q(N,Z,B) \geq 
E_{\rm conf,hf}(N,Z,B)(1-C_1 {\mathcal R}_2),
\end{align}
where
\begin{align}
\label{eq:R2}
{\mathcal R}_2 = \delta^{-1}  \frac{Z}{\sqrt{B}} (\delta L B^{1/2})^{\mu} \times \begin{cases} \beta^{-3/5}, & B \leq 2Z^3 \\
\min(B^{-1/3}, \frac{Z}{\sqrt{B}} \log \frac{B}{Z^3} ),& B\geq 2Z^3.
\end{cases}
\end{align}
\end{thm}

Theorem~\ref{thm:BiggerConf} implies a bound on the perpendicular kinetic energy of $P_>^N \psi$.

\begin{cor}
\label{cor:Biggerkinetic}~\\
Let the hypothesis of Theorem~\ref{thm:BiggerConf} be satisfies and let $\psi = \psi_{N,Z,B}$ be a normalised ground state for $H(N,Z,B)$.
Then
the perpendicular kinetic energy of $P_>^N \psi$ satisfies the estimate
\begin{eqnarray}
\label{eq:KhatR2}
   0 \leq \langle \psi \,,\, \hat{K}^N P_>^N \psi \rangle \leq C_1 {\mathcal R}_2
{\mathcal E}(Z,B),
\end{eqnarray}
where ${\mathcal R}_2$ is defined by \eqref{eq:R2}.
\end{cor}

\begin{proof}[Proof of Corollary~\ref{cor:Biggerkinetic}]~\\
Define $\tilde{H}_{\frac{1}{2}}(N,Z,B) := H(N,Z,B) - \frac{1}{2}\hat{K}^N P_>^N$ and proceed as in the proof of Corollary~\ref{cor:kinetic}.
\end{proof}

\begin{proof}[Proof of Theorem~\ref{thm:BiggerConf}]~\\
The first inequality in \eqref{eq:NewConf} is an easy consequence of the variational principle, so we only need to prove the second.

We define, for $\alpha \subseteq \{1, \ldots, N\}$, $\tilde{\alpha} =  \{1, \ldots, N\} \setminus \alpha$ and 
\begin{align}
\label{eq:Palpha}
P^{\alpha} := \prod_{j \in \alpha} P_0^{(j)}  \prod_{k \in \tilde{\alpha}} P_>^{(k)} 
\end{align}
We get the identities
\begin{align}
\sum_{\alpha \subseteq \{1, \ldots, N\}} P^{\alpha} = 1, \quad\quad\text{and}\quad\quad
P^{\alpha} P^{\alpha'} = 0 \, \text { for } \, \alpha \neq \alpha'.
\end{align}
Proceeding as in the proof of \eqref{eq:decomp_new2}, 
we get the following operator inequality (where we write $r_{jk}$ instead of  $x^{(j)}-x^{(k)}$):
\begin{align}
\label{eq:SplitQ}
H \geq \sum_{\alpha} P^{\alpha} \Big( \hat{Q}^{\alpha} + \tilde{Q}^{\alpha} \Big) P^{\alpha},
\end{align}
with
\begin{align}
\label{eq:DefQhat}
\hat{Q}^{\alpha}&= \sum_{j \in \alpha} \Big\{H_{\bf A}^{(j)} - Z V(x^{(j)})
- \epsilon Z \Pi_>^{(j)} V(x^{(j)}) \Pi_>^{(j)}-\epsilon B^{1/2} Z \Pi_0^{(j)} W_1(x^{(j)}) \Pi_0^{(j)}\nonumber\\
&\quad\quad
+\sum_{k\neq j} \Big( \frac{1}{2} V(r_{jk}) - \frac{3}{2}\epsilon \Pi_>^{(j)} V(r_{jk}) \Pi_>^{(j)}
- \frac{3}{2}\epsilon B^{1/2}\Pi_0^{(j)} W_1(r_{jk}) \Pi_0^{(j)}\Big)\Big\},
\intertext{ and }
\label{eq:DefQtilde}
\tilde{Q}^{\alpha} &=\sum_{j \in \tilde{\alpha}}\Big\{
H_{{\bf A}}^{(j)} - Z V(x^{(j)})
- \epsilon^{-1} Z  V(x^{(j)})-\epsilon^{-1} B^{1/2} Z  W_2(x^{(j)}) \nonumber\\
&\quad\quad\quad\quad
+ \sum_{k \neq j} \Big(\frac{1}{2r_{jk}}
- \frac{3}{2} \epsilon^{-1} (B^{1/2}W_2(r_{jk})  +V(r_{jk}) \Big)
\Big\}.
\end{align}

Notice right away that, since we will choose $\epsilon \ll1$
and $|W_2(x)| \leq \frac{2{\mathcal C}}{B^{1/2}|x|}$ (and ${\mathcal C}\geq 1$), $\tilde{Q}^{\alpha}$ satisfies
\begin{align}
\label{eq:BoundTildeQ}
\tilde{Q}^{\alpha} \geq \sum_{j \in \tilde{\alpha}}\Big\{
H_{{\bf A}}^{(j)} - 4{\mathcal C}\epsilon^{-1} Z  V(x^{(j)}) - 6{\mathcal C} \epsilon^{-1} \sum_{k \neq j} V(r_{jk})\Big\}.
\end{align}

\noindent{{\bf Bound on $\tilde{Q}$}.}\\
We will prove that for all $\mu \in (0,1/2)$, there exists $c_0>0$ such that if
$\delta, \epsilon$ satisfy
\begin{align}
\label{eq:codnition2}
\frac{Z}{\epsilon \delta B L } (\delta L B^{1/2})^{\mu} \leq c_0,
\end{align}
then
\begin{align}
P^{\alpha} \tilde{Q}^{\alpha} P^{\alpha}  \geq \frac{1}{8} B P^{\alpha}.
\end{align}

We rewrite \eqref{eq:BoundTildeQ} as
\begin{align*}
\tilde{Q}^{\alpha} \geq \sum_{j \in \tilde{\alpha}} \Big\{
\frac{1}{2} H_{{\bf A}}^{(j)} &- 4 {\mathcal C} \epsilon^{-1} Z V(x^{(j)})\\
&+
\frac{1}{N-1}\sum_{k \neq j}\Big( \frac{1}{2} H_{{\bf A}}^{(j)} - 4 {\mathcal C} \epsilon^{-1} (N-1) V(x^{(j)})\Big)
\Big\}.
\end{align*}
It thus suffices to prove that if \eqref{eq:codnition2} is satisfied then, for all $z \in {\mathbb R}^3$,
\begin{align}
\label{eq:repeatLemma}
P_>^{(j)} \Big\{ \frac{1}{2} H_{{\bf A}}^{(j)} &- 4 {\mathcal C} \epsilon^{-1} Z V(x^{(j)}-z) \Big\} P_>^{(j)}
\geq \frac{B}{8} P_>^{(j)} .
\end{align}
But \eqref{eq:repeatLemma} is exactly the result of Lemma~\ref{lem:IV} below.

To assure the condition \eqref{eq:codnition2}, we choose\footnote{
We can rewrite the choice of $\epsilon$ as
$\epsilon = \frac{Z}{B^{1/2}} (\delta L B^{1/2})^{\mu -1} M$. 
Therefore the facts that (see \eqref{eq:deltaLower} and \eqref{eq:CondParam}) $\delta \gg \frac{B^{-1/2}}{Z}$ and $Z^2 \ll B$ imply that the previously used assumption, $\epsilon \ll 1$, is satisfied.
}
\begin{align}
\label{eq:CondEpsilon}
\epsilon= \frac{Z}{\delta B L} (\delta L B^{1/2})^{\mu} M.
\end{align}
for some large constant $M$.

\noindent{{\bf Bound on $\hat{Q}$}.}\\
Throwing away the positive terms $r_{jk}^{-1}$ with $j,k \in \tilde{\alpha}$, we can estimate, for some $\eta>0$, $\hat{Q}$ as
\begin{align}
\hat{Q} &\geq (1-3\eta)\Big(\sum_{j\in \alpha} (H_{{\bf A}}^{(j)} - ZV(x^{(j)})) + \sum_{j,k \in \alpha, j<k} r_{jk}^{-1}\Big) \nonumber\\
&\quad\quad+\eta(\hat{Q}_1+\hat{Q}_2+\hat{Q}_3),
\end{align}
with
\begin{align*}
\hat{Q}_1&=\sum_{j\in \alpha} H_{{\bf A}}^{(j)} - 3ZV(x^{(j)}) + 
\sum_{j,k \in \alpha, j<k} V(r_{jk}),\\
\hat{Q}_2&=\sum_{j\in \alpha} \Big\{H_{{\bf A}}^{(j)}-\epsilon \eta^{-1}B^{1/2} Z \Pi_0^{(j)} W_1(x^{(j)}) \Pi_0^{(j)}
- \frac{3\epsilon}{2\eta} B^{\frac{1}{2}}\sum_{k \neq j}\Pi_0^{(j)} W_1(r_{jk}) \Pi_0^{(j)}\Big\},
\end{align*}
and
\begin{align*}
\hat{Q}_3&=\sum_{j\in \alpha} \Big\{H_{{\bf A}}^{(j)}
- \epsilon \eta^{-1}Z \Pi_>^{(j)} V(x^{(j)}) \Pi_>^{(j)}
- \frac{3\epsilon}{2\eta} \sum_{k \neq j}\Pi_>^{(j)} V(r_{jk}) \Pi_>^{(j)}\Big\}.
\end{align*}

To estimate the operators $Q_j$'s, we will use the same strategy as always: First we use the L\'{e}vy-Leblond formula as in \eqref{eq:levyLe} to effectively have to estimate one-body operators, then we use a suitable Lieb-Thirring inequality---choosing the optimal one in each case from Proposition~\ref{prop:LT}.

The term $\hat{Q}_2$ is estimated exactly as in Section~\ref{LocPi0}---the $\Pi_0^{(j)}$'s surrounding the potential assuring that we can use the Lieb-Thirring inequality from the lowest Landau band, Proposition~\ref{prop:LT}(\ref{LT-mag_lowest})---and we get
\begin{align}
\label{eq:Q23}
P^{\alpha} \hat{Q}_2 P^{\alpha} \geq -c \eta^{-3/2} \epsilon^{3/2} Z^{3/2} B^{1/4}.
\end{align}

Notice that $\hat{Q}_1$ is an atomic Pauli Hamiltonian for $|\alpha|$ electrons and nuclear charge $3Z$.  We can therefore estimate
\begin{align}
\label{eq:Q1}
\hat{Q}_1 \geq E^{\rm Q}(|\alpha|, 3Z, B) \geq E^{\rm Q}(N, 3Z, B)
\geq -C {\mathcal E}(N,Z,B).
\end{align}

Finally we consider $\hat{Q}_3$. Proceeding as for $\hat{Q}_2$ but using the standard magnetic Lieb-Thirring estimate, Proposition~\ref{prop:LT}(\ref{LT-classic}), we end up with 
\begin{align*}
P^{\alpha}\hat{Q}_3P^{\alpha}&\geq -C \int_{{\mathbb R}^3} \Big[B - \frac{\epsilon Z}{\eta |x|}\Big]_{-}^{5/2}\,dx 
= -C B^{5/2} \int_0^1 [1 - \rho^{-1}]^{5/2} \Big(\frac{\epsilon Z}{B\eta}\Big)^3 \rho^2\,d\rho \\
&= -C'\epsilon^3 \eta^{-3} Z^3 B^{-1/2}.
\end{align*}

\noindent {\bf Choice of parameters.} $B \leq 2Z^3$.\\
Recall that here 
$$
L = Z^{-2/5} B^{-1/5}, \quad\quad
\frac{B^{-1/2}}{L} = \beta^{-3/10}.
$$
Write, for some $\tilde{R}>0$,
\begin{align*}
\eta = \tilde{R} \beta^{-3/5} = \tilde{R} B^{-3/5} Z^{4/5},
\end{align*}
and 
\begin{align}
\label{eq:SomeEpsilon}
\epsilon = R \frac{Z}{\sqrt{B}}\quad \text{ with } R:=\delta^{-1}\frac{B^{-1/2}}{ L} (\delta L B^{1/2})^{\mu} M.
\end{align}
Then we find
\begin{align}
\label{eq:Q3}
P^{\alpha}( \hat{Q}_1+\hat{Q}_2&+\hat{Q}_3) P^{\alpha}\nonumber\\
&\geq
-C \Big\{1+\Big(\frac{R}{\tilde{R}}\Big)^{3/2}   + \Big(\frac{R}{\tilde{R}}\Big)^3 \Big(\frac{Z^3}{B}\Big)^{3/5} \Big\}B^{2/5} Z^{9/5}P^{\alpha}.
\end{align}
We choose
\begin{align}
\tilde{R}= R \Big(\frac{Z^3}{B}\Big)^{1/5}= \delta^{-1} \frac{Z}{\sqrt{B}}  (\delta L B^{1/2})^{\mu} M. 
\end{align}
Then 
$$
\Big(\frac{R}{\tilde{R}}\Big)^3 \Big(\frac{Z^3}{B}\Big)^{3/5}=1,\quad\quad
\Big(\frac{R}{\tilde{R}}\Big)^{3/2} = \Big(\frac{Z^3}{B}\Big)^{-3/10} \leq 2^{3/10},
$$
and our final estimate becomes
\begin{align}
\label{eq:old427}
P^{\alpha}\big( \hat{Q}^{\alpha}+\tilde{Q}^{\alpha}\big)
P^{\alpha}\geq
P^{\alpha}\big(  E_{\rm conf,hf} -C \delta^{-1} \frac{Z}{\sqrt{B}} (\delta L B^{1/2})^{\mu} B^{-1/5} Z^{13/5}\big).
\end{align}
By \eqref{eq:SplitQ} and \eqref{eq:old427} we get
\begin{align}
E(N,Z,B) \geq E_{\rm conf,hf} -C \delta^{-1} \frac{Z}{\sqrt{B}} (\delta L B^{1/2})^{\mu} B^{-1/5} Z^{13/5}.
\end{align}
This finishes the proof of Theorem~\ref{thm:BiggerConf} in the case $B \leq 2Z^3$.

~

\noindent{\bf Choice of parameters.} $B\geq 2Z^3$.\\
Here $L= Z^{-1} [\log \frac{B}{Z^3}]^{-1}$.
First we use the same approach as for the case $B\leq 2Z^3$. Then we get the estimate
\begin{align}
P^{\alpha} \hat{Q} P^{\alpha} \geq
\Big\{&(1-3\eta) E_{\rm conf,hf}\nonumber\\
&-C\eta\big[1 + (\frac{\epsilon}{\eta})^3B^{-1/2}(\log \frac{B}{Z^3})^{-2} \nonumber\\
&\quad\quad\quad+ (\frac{\epsilon}{\eta})^{3/2}
B^{1/4} Z^{-3/2}(\log \frac{B}{Z^3})^{-2} \big] Z^3 (\log \frac{B}{Z^3})^2 \Big\} P^{\alpha}. \nonumber
\end{align}
This follows from \eqref{eq:Q23}, \eqref{eq:Q1}, and \eqref{eq:Q3}. 

We can now choose $\epsilon$ as in \eqref{eq:CondEpsilon} and
\begin{align}
\eta =  \big( \frac{B}{Z^6} \big)^{1/6} \epsilon = \delta^{-1} \big( \frac{B}{Z^6} \big)^{1/6}  \frac{Z^2 \log \frac{B}{Z^3}}{B} (\delta L B^{1/2})^{\mu} M.
\end{align}
Then
\begin{align*}
&(\frac{\epsilon}{\eta})^{3/2}
B^{1/4} Z^{-3/2}(\log \frac{B}{Z^3})^{-2}  = (\log \frac{B}{Z^3})^{-2} \leq (\log 2)^{-2}, \\
&(\frac{\epsilon}{\eta})^3B^{-1/2} = \frac{Z^3}{B} \leq 1/2.
\end{align*}
This leads to the first error bound in \eqref{eq:R2} for $B \geq 2Z^3$.

To get the second estimate in \eqref{eq:R2} for $B \geq 2Z^3$ we estimate $\hat{Q}_2$ and $\hat{Q}_3$ a bit differently.
We take
$$
\eta = \epsilon = \delta^{-1} \frac{Z^2 \log \frac{B}{Z^3}}{B} (\delta L B^{1/2})^{\mu} M .
$$
The argument applied to $\tilde{Q}$ gives that $\hat{Q}_3 \geq 0$. We will use Theorem~\ref{thm:Hydrogen} to estimate $\hat{Q}_2$. To prepare for this we write
\begin{align}
\hat{Q}_2 \geq \frac{1}{2}\sum_{j \in \alpha} \Pi_0^{(j)}\Big\{&
H_{\bf A}^{(j)} - 2B^{1/2} Z  W_1(x^{(j)}) \nonumber\\
&+
\frac{1}{N-1} \sum_{k \neq j} \big( H_{\bf A}^{(j)} - 3B^{1/2}(N-1) W_1(r_{jk}) 
\big)\Big\} \Pi_0^{(j)}.
\end{align}
We now estimate each $W_1(z) \leq \frac{2{\mathcal C}}{B^{1/2} |z|}$, and apply Theorem~\ref{thm:Hydrogen} with $N=1$ to each of the $N$ operators inside the $\{ \cdot \}$.
Counting terms and using $N \approx Z$, we get
$$
\hat{Q}_2 \geq -C Z^3 [\log \frac{B}{Z^3}]^2.
$$
This finishes the proof of the second error bound in \eqref{eq:R2} for $B \geq 2Z^3$ and therefore of Theorem~\ref{thm:BiggerConf}.
\end{proof}

\section{The current}
\label{current}
\subsection{Discussion}~\\
For reasons that will become clear later, we will generally impose the technical restriction on the test functions ${\bf a}$ in \eqref{eq:CurrDef}, \eqref{eq:CurrOp} that they be everywhere perpendicular to the magnetic field, i.e.
\begin{align}
\label{eq:restriction}
{\bf a} = (a_1, a_2, 0).
\end{align}
The identity \eqref{eq:curVan} below is valid under this assumption and \eqref{eq:restriction} is also a crucial hypothesis for the validity of Theorem~\ref{thm:Commutator}. The missing third component of the current can be reconstructed from the remaining two using the spatial symmetries of the Hamiltonian and gauge invariance. So our final result on the current, Theorem~\ref{thm:summarize_improved}, does {\it not} suppose \eqref{eq:restriction}.

In the regime where $B$ is large compared to $Z^{4/3}$, the ground state $\psi$ is essentially, 
in the sense of Theorem~\ref{thm:loc_improved}, localised to the lowest Landau band, $\Ran \Pi_0^N$ (with the notation from \eqref{eq:PiN}).
The energy and the density can be correctly calculated to leading order by restricting to this subspace, but the current satisfies
\begin{align}
\label{eq:curVan}
\Pi_0^N J({\bf a}) \Pi_0^N = 0.
\end{align}
Thus, one needs more than leading order information on $\psi$ to calculate the current.

One can consider the current operator $J({\bf a})$ as composed of two contributions, a {\it spin-current} $B \sum {\bf b} \cdot \sigma$ and a {\it persistent current}. The identity \eqref{eq:curVan} expresses that the sum of these operators cancel on the lowest Landau band. If we were to calculate the separate contributions of these two terms to the total current, the analysis would be much easier.

\subsection{Splitting the current}
\label{previous}~\\
A first step towards the calculation of the current is to replace the operator $J({\bf a})$ by another operator having the same matrix element in the ground state $\psi$ and being easier to analyse.
This was realised in \cite{fournais4} and the result is given in \eqref{commutator_atomer} below.

\begin{thm}
\label{thm:Commutator}~\\
Let ${\bf a}=(a_1,a_2,0)\in C_0^{\infty}$ be given and define
$$
{\bf \tilde{a}}=(-a_2,a_1,0), \quad\quad{\bf \tilde{a}_0}(x)= {\bf \tilde{a}}(x)-{\bf \tilde{a}}(0).
$$
Define furthermore $M_{\bf a}$ as the negative, symmetrised Jacobian matrix of ${\bf \tilde{a}}$,
$$
M_{\bf a} = -\big( D{\bf \tilde{a}} + (D{\bf \tilde{a}})^t\big)
= \begin{pmatrix}
2\partial_1 a_2 & \partial_2 a_2-\partial_1 a_1 &\partial_3 a_2\\
\partial_2 a_2-\partial_1 a_1 & -2\partial_2 a_1 & -\partial_3 a_1 \\
\partial_3 a_2 & -\partial_3 a_1 & 0
\end{pmatrix}.
$$
Define finally a decomposition of the Laplacian,
$$
\Delta_{\perp} := \frac{\partial^2}{\partial^2_1} + \frac{\partial^2}{\partial^2_2}, \quad\quad
\Delta_{\parallel} := \frac{\partial^2}{\partial^2_3}, 
$$
and the operators
\begin{align}
\label{eq:Jkin}
J_{{\rm KIN}} &=
\sum_{j=1}^N {\bf p}_{{\bf A}}^{(j)}
M_{\bf a}(x^{(j)}){\bf
p}_{{\bf A}}^{(j)} + B \sigma
\cdot {\bf b}(x^{(j)}) -
\tfrac{1}{2} \Delta_{\perp} b_3(x^{(j)}),
\displaybreak[0]\\
\label{eq:Jdens}
J_{{\rm DENS}} &= \sum_{j=1}^N \left( Z \frac{{\bf
\tilde{a}_0}(x^{(j)})\cdot x^{(j)}}{|x^{(j)}|^3} -
\tfrac{1}{2} \Delta_{\parallel} b_3(x^{(j)}) \right),
\displaybreak[0]\\
\label{eq:Jint}
J_{{\rm INT}} &=
\sum_{1\leq j< k\leq N} \frac{(x^{(j)} - x^{(k)})\cdot (
{\bf \tilde{a}}(x^{(j)}) - {\bf \tilde{a}}(x^{(k)}))}{|x^{(j)} - x^{(k)} |^3}\;.
\end{align}
Then, for any eigenfunction $\psi$ of $H(N,Z,B)$, we have the identity
\begin{align}
\label{commutator_atomer}
\langle \psi, J({\bf a}) \psi \rangle
=
\big\langle \psi, (J_{{\rm KIN}} - J_{{\rm INT}} + J_{{\rm
DENS}}) \psi \big\rangle.
\end{align}
\end{thm}

\begin{remark}~\\
The splitting of the (lower order) term $\Delta b_3$ between $J_{{\rm DENS}}$ and $J_{{\rm KIN}}$ may seem a bit arbitrary at this point. The reason for including a part of this term in $J_{{\rm KIN}}$ is that it is convenient to have the identity \eqref{eq:J1-vanish} below. Furthermore, if one were to extend the analysis of the current to magnetic field strengths above $Z^3$, this splitting seems to be the natural one, since in that case $\Delta_{\perp} b_3$ is no longer of lower order.
\end{remark}

The identity \eqref{commutator_atomer} is valid for any eigenstate $\psi$ for $H(N,Z,B)$, in particular for the ground state.
It comes from expressing $J({\bf a})- (J_{{\rm KIN}} - J_{{\rm INT}} + J_{{\rm
DENS}})$ as a commutator $i[H(N,Z,B), O]$, where the operator $O$ can be chosen as 
$$
O = \frac{i}{2}\sum_{j=1}^N \Big\{ {\bf \tilde{a}_0}(x^{(j)}) \cdot {\bf p}_{{\bf A}}^{(j)}+{\bf p}_{{\bf A}}^{(j)}\cdot {\bf \tilde{a}_0}(x^{(j)})
\Big\}.
$$
The commutator vanishes in an eigenstate,
$$
\langle \psi,  [H(N,Z,B), O] \psi \rangle =0,
$$ 
and \eqref{commutator_atomer} follows (see \cite{fournais4} for details).
This sketch of a proof gives an idea why the construction in \cite{Fournais3} of approximate eigenstates with `wrong' current works: By perturbing the eigenstate $\psi$ a little bit, one can get an enormous contribution from the commutator $\langle \psi, [H(N,Z,B), O] \psi \rangle$ without changing the energy $\langle \psi, H \psi \rangle$ very much.

The restriction on ${\bf a}$ from \eqref{eq:restriction} is forced upon us by the use of the formula \eqref{commutator_atomer}. Only for ${\bf a} \cdot {\bf B}=0$ is it possible to find an ${\bf \tilde{a}}$ corresponding to ${\bf a}$.

The right side of \eqref{commutator_atomer} is much easier to analyse than the left side. The operator $J_{{\rm DENS}} = \sum_{j=1}^N \phi(x^{(j)})$, with $\phi(x)=\frac{{\bf \tilde{a}_0}(x)\cdot x}{|x|^3} -
\tfrac{1}{2} \Delta_{\parallel} b_3(x)$, is just a sum of one-particle multiplication operators, and therefore
\begin{align}
\label{eq:density}
\langle \psi,  J_{{\rm DENS}} \psi \rangle = \int_{{\mathbb R}^3} \phi \rho \,dx.
\end{align}
The operator $J_{{\rm INT}}=\sum\frac{(x^{(j)} - x^{(k)})\cdot (
{\bf \tilde{a}}(x^{(j)}) - {\bf \tilde{a}}(x^{(k)}))}{|x^{(j)} - x^{(k)} |^3}$ is similar to the interelectronic repulsion $\sum |x^{(j)}-x^{(k)}|^{-1}$. It has been analysed and calculated to leading order in \cite{fournais4,fournais8}.

In the following sections, we will calculate each of the terms, $\langle \psi , J_{{\rm DENS}} \psi \rangle$, 
$\langle \psi , J_{{\rm INT}} \psi \rangle$ and $\langle \psi , J_{{\rm KIN}} \psi \rangle$ independently in terms of magnetic Thomas-Fermi theory. Before we do so, let us notice how those results will imply Theorem~\ref{thm:summarize_improved}.

\begin{proof}[Proof of Theorem~\ref{thm:summarize_improved}]~\\
Using the symmetry \eqref{eq:Symmetry} of $\psi$ it suffices to consider ${\bf a}$ where $a_3$ is odd in $x_3$:
$$
a_3(x_1,x_2,-x_3) = -a_3(x_1,x_2,x_3).
$$
But such an $(0,0,a_3)$ with $a_3$ odd in $x_3$ is gauge equivalent to a compactly supported $(a_1,a_2,0)$. Thus it suffices to prove Theorem~\ref{thm:summarize_improved} in the case where \eqref{eq:restriction} is satisfied, i.e. $a_3=0$. In that case we can use the formula \eqref{commutator_atomer} to replace the operator $J({\bf a})$ by the three operators $J_{{\rm KIN}}, J_{{\rm INT}}$ and $J_{{\rm DENS}}$.

By Theorems~\ref{thm:CalcInMTF} and~\ref{thm:VirialMTF} below we can write 
\begin{align}
\frac{d}{dt} {\Big |}_{t=0} E^{{\rm MTF}}(N,Z,B{\bf e}_z+t B \curl {\bf
a})  = J^{\rm MTF}_{\rm KIN} - J^{\rm MTF}_{\rm INT}+ J^{\rm MTF}_{\rm DENS},
\end{align}
where the right hand side is defined in \eqref{eq:J-MTF}. In order to prove Theorem~\ref{thm:summarize_improved} it thus suffices to prove the three estimates
\begin{align}
\langle \psi,  J_{{\rm KIN}} \psi \rangle &= J^{\rm MTF}_{\rm KIN} + o({\mathcal E}(Z,B)), \nonumber\\
\langle \psi,  J_{{\rm INT}} \psi \rangle &= J^{\rm MTF}_{\rm INT} + o({\mathcal E}(Z,B)), \nonumber\\
\langle \psi,  J_{{\rm DENS}} \psi \rangle &= J^{\rm MTF}_{\rm DENS} + o({\mathcal E}(Z,B)).
\end{align}
These three estimates are the results of Theorems~\ref{thm:Jkin}, \ref{thm:Jint} and \ref{thm:Jdens} below respectively. This reduces the proof of Theorem~\ref{thm:summarize_improved} to the proof of those three theorems.
\end{proof}

\section{Current in MTF-theory}
\label{CurrentMTF}
In this section we briefly recall results on the MTF-functional. Details can be found in \cite{LSY} and \cite{fournais4,fournais8} (see also \cite{Lieb-Simon,Lieb} for general theory of Thomas-Fermi-type models).
We start by considering general magnetic fields, ${\bf B} \in L^{\infty}_{{\rm loc}}({\mathbb R}^3)$. With ${\mathcal E}_{Z, {\bf B}}^{\rm MTF}$ as defined in \eqref{eq:E-MTF} on the domain 
${\mathcal C}_{N,{\bf B}}$ from \eqref{eq:DefCB} we have the following result.

\begin{thm}
\label{thm:CalcInMTF}~
\begin{itemize}
\item
There exists a unique $\rho^{\rm MTF} = \rho^{\rm MTF}_{{\bf B},N,Z} \in {\mathcal C}_{N,{\bf B}}$ such that 
$$
{\mathcal E}_{Z, {\bf B}}^{\rm MTF}[\rho^{\rm MTF}] = E^{{\rm MTF}}(N,Z,{\bf B}).
$$
Furthermore, there exists a critical particle number $N_c=N_c(Z,{\bf B})>0$ such that
$\int \rho^{\rm MTF}_{{\bf B},N,Z} \,dx = \min(N,N_c)$.
\item The minimizer satisfies the Thomas-Fermi equation
\begin{align}
\label{eq:TFeq}
\tau'_{|{\bf B}|}(\rho^{\rm MTF}) = \big[ V_Z(x) + \rho^{\rm MTF}*|x|^{-1} + \mu \big]_{-},
\end{align}
for some unique (chemical potential) $\mu = \mu(N,Z,{\bf B})$ such that $\mu(N,Z,{\bf B}) = 0$ for $N>N_c$. With the definition $V_{{\rm eff}}:= V_Z(x) + \rho^{\rm MTF}*|x|^{-1} + \mu$, we can rewrite \eqref{eq:TFeq} as 
\begin{align}
\rho^{\rm MTF} = P_{[{\bf B}|}'(|V_{{\rm eff}}]_{-}),
\end{align}
and
\begin{align}
\label{eq:TFeq3}
P_{|{\bf B}|}([V_{{\rm eff}}]_{-}) = - \tau_{|{\bf B}|}(\rho^{\rm MTF})+ \rho^{\rm MTF} [V_{{\rm eff}}]_{-}.
\end{align}
\item 
Suppose that $|{\bf B}| \geq c >0$.
Then, for all ${\bf a} \in C_0^{\infty}({\mathbb R}^3, {\mathbb R}^3)$, the map $t \mapsto E^{{\rm MTF}}(N,Z,{\bf B}+t \, \curl {\bf a})$ is differentiable at $t=0$ and defines a current ${\bf j}^{\rm MTF}$ by
\begin{align}
\label{eq:MTFcurrent}
\int {\bf j}^{\rm MTF} \cdot {\bf a}\,dx &:= \frac{d}{dt} E^{{\rm MTF}}(N,Z,{\bf B}+t \, \curl {\bf a}) \big |_{t=0} \nonumber\\
&=\int \frac{{\bf B} \cdot {\bf b}}{|{\bf B}|^2} \big\{ [V_{{\rm eff}}]_{-} P_{|{\bf B}|}'([V_{{\rm eff}}]_{-}) - \frac{5}{2} P_{|{\bf B}|}([V_{{\rm eff}}]_{-})\big\}\,dx,
\end{align}
where ${\bf b} = (b_1, b_2, b_3)= \curl {\bf a}$.
\end{itemize}
\end{thm}

We now restrict ourselves to the case of a constant magnetic field ${\bf B} = (0,0,B)$.
Notice from \eqref{eq:MTFcurrent} that ${\bf j}^{\rm MTF} \perp {\bf B}$, i.e. it suffices to consider test vector fields ${\bf a}$ of the form ${\bf a} = (a_1, a_2, 0)$.

For large field strength  the formula \eqref{eq:MTFcurrent} is not convenient for comparison with the quantum current. In order to transform the expression we consider ${\bf a} = (a_1, a_2, 0)$ and define
$$
{\bf \tilde{a}} = (-a_2,a_1,0), \quad
{\bf \tilde{a}_0}(x) =  {\bf \tilde{a}}(x) - {\bf \tilde{a}}(0).
$$
Let furthermore, $\rho^{\rm MTF}$ be the minimizer of ${\mathcal E}_{Z, {\bf B}}^{\rm MTF}$ and define (for small $t$)
$$
\Lambda_t(x) = \det (I + t D {\bf \tilde{a}_0}(x)), \quad
\rho_t(x) = \Lambda_t(x) \rho^{\rm MTF}(x+t {\bf \tilde{a}_0}(x)).
$$
Notice that $\Lambda_t = 1 + t\,\tr[ D {\bf \tilde{a}_0}] + {\mathcal O}(t^2) = 1 - t b_3 + {\mathcal O}(t^2)$.
Define finally, the diffeomorphism $\phi_t$ by $\phi_t(x+t{\bf \tilde{a}_0}(x))= x$.
Changing variables in the integrals we can calculate
\begin{align}
{\mathcal E}_{Z, {\bf B}}^{\rm MTF}[\rho_t] &=
\int \tau_B(\Lambda_t(\phi_t(y)))\frac{dy}{\Lambda_t(\phi_t(y))} \nonumber \\
&\quad-
\int \frac{Z}{|\phi_t(y)|} \rho(y)\,dy
+ \frac{1}{2} \iint \frac{\rho(x)\rho(y)}{|\phi_t(x)-\phi_t(y)|}\,dxdy.
\end{align}
Using that the derivative of ${\mathcal E}_{Z, {\bf B}}^{\rm MTF}[\rho_t]$ at $t=0$ has to vanish, combined with the Thomas-Fermi equation \eqref{eq:TFeq3}, we get the relation
\begin{align}
\int b_3 P_{B}(|V_{{\rm eff}}|_{-}) = D_{{\bf \tilde{a}}}(\rho^{\rm MTF}, \rho^{\rm MTF})-
\int \frac{Z x \cdot {\bf \tilde{a}_0}(x)}{|x|^3} \rho^{\rm MTF}(x)\,dx,
\end{align}
where 
\begin{align*}
D_{{\bf \tilde{a}}}(f,g) := \frac{1}{2} \iint \overline{f(x)} \frac{(x-y)\cdot\big({\bf \tilde{a}}(x)-{\bf \tilde{a}}(y)\big)}{|x-y|^3} g(y)\,dxdy.
\end{align*}
Therefore we can write the MTF-current in the case of constant magnetic field as
\begin{align}
\label{eq:SplittingMTF}
\int {\bf j}^{\rm MTF} \cdot B {\bf a}\,dx = J^{\rm MTF}_{\rm KIN} - J^{\rm MTF}_{\rm INT}+ J^{\rm MTF}_{\rm DENS},
\end{align}
where
\begin{align}
\label{eq:J-MTF}
J^{\rm MTF}_{\rm KIN} &= \int b_3 \big\{ [V_{{\rm eff}}]_{-} P_{B}'([V_{{\rm eff}}]_{-}) - \frac{3}{2} P_{B}(|V_{{\rm eff}}|_{-})\big\}\,dx,\nonumber\\
J^{\rm MTF}_{\rm INT} &= D_{{\bf \tilde{a}}}(\rho^{\rm MTF}, \rho^{\rm MTF}),\nonumber\\
J^{\rm MTF}_{\rm DENS} &= \int \frac{Z x \cdot {\bf \tilde{a}_0}(x)}{|x|^3} \rho^{\rm MTF}(x)\,dx.
\end{align}

\begin{remark}~\\
For weak magnetic fields the discrete sum in the definition of $P_B$ (see \eqref{eq:Pb}) can be approximated by the corresponding integral and one finds $P_B(v) \approx const \times v^{5/2}$. On the other hand, if $B$ is strong then only the first term in the sum contributes, and $P_B(v) \approx \frac{B}{3 \pi^2} v^{3/2}$. For the atomic MTF-problems these approximations are correct to leading order if $B \ll Z^{4/3}$ (weak field) and $B \gg Z^{4/3}$ (strong field). The original formula for the current, \eqref{eq:MTFcurrent}, thus suggests (and a rigorous analysis confirms) that
$$
\int {\bf j}^{\rm MTF} \cdot B {\bf a}\,dx = o(E^{\rm MTF}(N,Z,B)),
$$
for $B \ll Z^{4/3}$. Similarly, the changed homogeneity for large fields gives that
$$
J^{\rm MTF}_{\rm KIN} = o(E^{\rm MTF}(N,Z,B)),
$$
for $B \gg Z^{4/3}$.
\end{remark}

\begin{thm}
\label{thm:VirialMTF}~\\
Let ${\bf B} = B(0,0,1)$ and $B,N,Z >0$. Then the MTF-current, ${\bf j}^{\rm MTF}$ defined in Theorem~\ref{thm:CalcInMTF} satisfies
\begin{itemize}
\item If ${\bf a} \parallel {\bf B}$, then $\int {\bf j}^{\rm MTF} \cdot {\bf a}\,dx = 0$.
\item If ${\bf a} \perp {\bf B}$, then ${\bf j}^{\rm MTF}$ satisfies \eqref{eq:SplittingMTF}.
\end{itemize}
Furthermore, for all ${\bf a}_{sc} \in C_0^{\infty}({\mathbb R}^3)$, all $\epsilon >0$ and all $\lambda > 0$ there exists a constant $C>0$ such that, with ${\bf a}(x) = \ell {\bf a}_{sc}(x/\ell)$, ($\ell$ being defined in \eqref{eq:ell}) if
$B \leq C^{-1} Z^{4/3}$, and $N/Z = \lambda$,
then
\begin{align}
\Big| \int {\bf j}^{\rm MTF} \cdot B {\bf a}\,dx \Big| \leq \epsilon {\mathcal E}(B,Z).
\end{align}
Also, if 
$B \geq C Z^{4/3}$, and $N/Z = \lambda$,
then
\begin{align}
\label{eq:KinMTFSmall}
\Big| J^{\rm MTF}_{\rm KIN} \Big| \leq \epsilon {\mathcal E}(B,Z).
\end{align}
\end{thm}

\section{Calculation of $J_{\rm DENS}$}
\label{Jdens}
This calculation was already carried out in \cite{LSY} but for completeness we give a sketch of the proof.

\begin{thm}
\label{thm:Jdens}~\\
Let ${\bf a_{sc}} = (a_{1,sc}, a_{2,sc},0) \in C_0^{\infty}({\mathbb R}^3)$ and define ${\bf a}(x) = \ell {\bf a_{sc}}(x/\ell)$ with $\ell$ from \eqref{eq:ell}. Let ${\bf \tilde{a}}, {\bf \tilde{a}_0}$ and ${\bf b}$ be as defined previously. Furthermore, let the operator $ J_{\rm DENS}$ be defined by \eqref{eq:Jdens} and let us define 
$$
W(x) = \frac{x\cdot {\bf \tilde{a}_0}(x)}{|x|^3}
$$
Let $\lambda>0$ and $\beta_{\infty} \in [0, +\infty]$ and let $(N, Z, B) =(N_n, Z_n, B_n)$ be a sequence 
with $Z \rightarrow \infty$, and 
such that 
$$
\lambda = N/Z, \quad B/Z^{4/3} \rightarrow \beta_{\infty}, \quad B/Z^3 \rightarrow 0,
$$
(as $n \rightarrow \infty$).
Then, if $\psi= \psi_{N,Z,B}$ is an associated sequence of ground states of $H(N,Z,B)$,
\begin{align}
\frac{1}{{\mathcal E}(Z,B)} \Big|\langle \psi, J_{\rm DENS} \psi \rangle - \int_{{\mathbb R}^3} Z W(x) \rho^{\rm MTF}_{B,N,Z}(x)\,dx \Big|
\rightarrow 0 \, .
\end{align}
\end{thm}

\begin{proof}[Sketch of proof of Theorem~\ref{thm:Jdens}]~\\
Since $\ell^{-2} \ll Z/\ell$ for $B\ll Z^3$, the term with $\Delta_{\parallel} b_3$ in $J_{\rm DENS}$ is clearly of lower order and will not be considered.

Let $M>0$ and consider $W_M(x) = 1_{\{ |x| \leq M\ell \}} W(x)$. For $M$ sufficiently big, we have 
$$
| W_M(x) - W(x) | \leq \frac{C}{M^2 \ell}\;,
$$
and therefore,
\begin{align}
\Big| \langle \psi, \sum_{j=1}^N Z\big( W_M(x^{(j)}) - W(x^{(j)})\big) \psi \rangle \Big| \leq
\frac{CZ}{M^2\ell} \int \rho\,dx \leq \frac{C'}{M^2} {\mathcal E}(Z,B).
\end{align}
A similar inequality holds on the Thomas-Fermi side and it therefore suffices to prove
\begin{align}
\label{eq:DensM}
\Big| Z \int \rho^{\rm Q}(x) W_M(x)\,dx - Z \int \rho^{\rm MTF}(x) W_M(x)\,dx\Big| = o({\mathcal E}(Z,B)),
\end{align}
for all $M>0$, where $\rho^Q$ is the density of the ground state $\psi$.

Consider, for $\alpha \in {\mathbb R}$, the self-adjoint operator (on the electronic Hilbert space ${\mathcal H}$)
$$
H_{\alpha}(N,Z,B) := H(N,Z,B) + \alpha Z \sum_{j=1}^N Z W_M(x^{(j)}),
$$
and define $E_{\alpha}(N,Z,B) := \inf \Spec H_{\alpha}(N,Z,B)$.

By the correspondence MTF---quantum mechanics proved in \cite{LSY}, we have for $\alpha =0$,
\begin{align}
\label{eq:UpperMTF}
E(N,Z,B) &= {\mathcal E}^{{\rm MTF}}_{Z,B}[\rho^{\rm MTF}] + o({\mathcal E}(Z,B)) \nonumber\\
&= -\int P_B([V_{\rm eff}]_{-})\,dx - D(\rho^{\rm MTF}, \rho^{\rm MTF}) - \mu N_c + o({\mathcal E}(Z,B)).
\end{align}
We need a similar lower bound on $E_{\alpha}(N,Z,B)$. The desired bound was already given in \cite{LSY}, but we recall the main line of reasoning for later reference.

~\\
\noindent{\it Lower bound on $E_{\alpha}(N,Z,B)$}\\
For $\phi \in \wedge_{j=1}^N L^2({\mathbb R}^3, {\mathbb C}^2)$ with $\| \phi \| = 1$ and density $\rho_{\phi}$, we estimate
\begin{align}
\langle \phi, H_{\alpha}(N,Z,B) \phi \rangle &\geq \inf \Spec {\mathfrak h}(N,B, V_{{\rm eff}}+\alpha W_M)
\nonumber\\
&+ \langle \phi, \sum_{j<k} \frac{1}{|x^{(j)}- x^{(k)}|} \phi \rangle - 2 D(\rho^{\rm MTF}, \rho_{\phi}) - \mu N,
\end{align}
where ${\mathfrak h}(N,B, V_{{\rm eff}}+\alpha W_M)$ is the mean field hamiltonian
\begin{align*}
&{\mathfrak h}(N,B, V_{{\rm eff}}+\alpha W_M) = \sum_{j=1}^N {\mathfrak h}^1(B, V_{{\rm eff}}+\alpha W_M)^{(j)},\\
& {\mathfrak h}^1(B, V_{{\rm eff}}+\alpha W_M) = p_{{\bf A}}^2 + {\bf B}\cdot \sigma + V_{{\rm eff}}+\alpha W_M.
\end{align*}
We now use the positivity of the Coulomb kernel
\begin{align}
\label{eq:DPos}
D(f,f) \geq 0, \quad \text{ for all }\quad f,
\end{align}
together with the Lieb-Oxford inequality, Theorem~\ref{thm:LO},
\begin{align}
\label{eq:LO}
\langle \phi, \sum_{j<k} \frac{1}{|x^{(j)}- x^{(k)}|} \phi \rangle \geq D(\rho_{\phi}, \rho_{\phi}) - C_{LO} \int \rho^{4/3}\,dx.
\end{align}
So we find
\begin{align}
\label{eq:BefSemiClas}
\langle \phi, H_{\alpha}(N,Z,B) \phi \rangle &\geq \inf \Spec {\mathfrak h}(N,B, V_{{\rm eff}}+\alpha W_M)
\nonumber\\
&- D(\rho^{\rm MTF}, \rho^{\rm MTF}) - \mu N 
- C_{LO} \int \rho_{\phi}^{4/3}\,dx.
\end{align}

At this point we need the semiclassical asymptotics of the mean field operator (cf. \cite[Theorem 3.1]{LSY}).

\begin{thm}[Magnetic semiclassics]\label{thm:MagnSemi}~\\
Suppose that $C>0$ and that $u_{B,Z}$ is a potential depending on the parameters $B,Z$ and such that with $h=\ell^{-1/2} Z^{-1/2}$, $b=B \ell^{3/2} Z^{-1/2}$:
\begin{itemize}
\item The quantity
$$ 
\frac{hb}{1+hb} \int [u_{B,Z}]_{-}^{3/2}\,dx + \frac{1}{1+hb} \int [u_{B,Z}]_{-}^{5/2}\,dx,
$$
is bounded uniformly for $B \leq C Z^3$.
\item For all $\epsilon>0$ there exists $R>0$ independent of $B,Z$ for $B\leq CZ^3$ such that
$$ 
\frac{hb}{1+hb} \int_{\{ |x|\geq R\}} [u_{B,Z}]_{-}^{3/2}\,dx + \frac{1}{1+hb} \int_{\{ |x|\geq R\}} [u_{B,Z}]_{-}^{5/2}\,dx< \epsilon.
$$
\end{itemize}
Then, with $U_{B,Z}(x) = Z\ell^{-1} u_{B,Z}(x/\ell)$ and $P_B$ being the pressure function introduced in \eqref{eq:Pb}, for all $\eta>0$ there exists $h_0>0$ such that if
$h<h_0$ then
\begin{align}
\Big| \frac{\tr[ {\mathfrak h}^{1}(B, U_{B,Z}) ]_{-}}{\int_{{\mathbb R^3}} P_B( [U_{B,Z}]_{-})} - 1 \Big | < \eta.
\end{align}
\end{thm}
Notice that when $Z \rightarrow \infty$, then $h \ll 1$ iff $B \ll Z^3$. Therefore, using Theorem~\ref{thm:MagnSemi} and the Thomas-Fermi equation \eqref{eq:TFeq3}, \eqref{eq:BefSemiClas} becomes
\begin{align}
\label{eq:mellem}
\langle \phi, H_{\alpha}(N,Z,B) \phi \rangle &\geq -(1+o(1)) \int P_B([V_{\rm eff} + \alpha W_M]_{-})\,dx
-D(\rho^{\rm MTF}, \rho^{\rm MTF})  \nonumber\\
&\quad - \mu N - C_{LO} \int \rho_{\phi}^{4/3}\,dx.
\end{align}
For $\phi$ such that $\langle \phi, H_{\alpha}(N,Z,B) \phi \rangle <0$ (the only ones where a lower bound is non-trivial) we can estimate
$$
\int \rho_{\phi}^{4/3}\,dx \leq C Z^{-2/3} {\mathcal E}(Z,B),
$$
by the Lieb-Thirring inequality (see \cite[p.121]{LSY} for details). The first term on the right hand side of \eqref{eq:mellem} has order of magnitude ${\mathcal E}(Z,B)$, so we can rewrite \eqref{eq:mellem} as
\begin{align}
\label{eq:Slut}
\langle \phi, H_{\alpha}(N,Z,B) \phi \rangle &\geq - \int P_B([V_{\rm eff} + \alpha W_M]_{-})\,dx
-D(\rho^{\rm MTF}, \rho^{\rm MTF})  \nonumber\\
&\quad\quad  - \mu N + o({\mathcal E}(Z,B)).
\end{align}
This is our lower bound on $E_{\alpha}(N,Z,B)$.

~\\
{\it Finishing the proof of Theorem~\ref{thm:Jdens}.}\\
We now combine \eqref{eq:UpperMTF} and \eqref{eq:Slut}.
The terms proportional to $\mu$ cancel, since $\mu=0$ for $N>N_c$.
Therefore, we can estimate as follows, using \eqref{eq:UpperMTF} and \eqref{eq:Slut} to get the last inequality
\begin{align}
\label{eq:Convex}
\alpha Z \int \rho^{\rm Q}(x)& W_M(x)\,dx =
\langle \psi, H_{\alpha}(N,Z,B) \psi\rangle - \langle \psi, H(N,Z,B) \psi\rangle \nonumber\\
&\geq E_{\alpha}(N,Z,B) - E(N,Z,B) \nonumber\\
&\geq \int P_B([V_{\rm eff}]_{-})- P_B([V_{\rm eff} + \alpha W_M]_{-})\,dx
+o({\mathcal E}(Z,B)).
\end{align}
Applying \eqref{eq:Convex} for positive and negative $\alpha$ we find\\
$\alpha >0$:
\begin{align}
\label{eq:alphaPos}
Z \int \rho^{\rm Q}(x) W_M(x)\,dx 
&\geq Z \int \frac{P_B([V_{\rm eff}]_{-})- P_B([V_{\rm eff} + \alpha W_M]_{-})}{\alpha}\,dx \nonumber\\
&\quad + \frac{1}{\alpha} o({\mathcal E}(Z,B)).
\end{align}
$\alpha <0$:
\begin{align}
\label{eq:alphaNeg}
Z \int \rho^{\rm Q}(x) W_M(x)\,dx
&\leq Z \int \frac{P_B([V_{\rm eff}]_{-})- P_B([V_{\rm eff} + \alpha W_M]_{-})}{\alpha}\,dx \nonumber\\
&\quad + \frac{1}{\alpha} o({\mathcal E}(Z,B)).\end{align}
Remembering the Thomas-Fermi equation $\rho^{\rm MTF} = P_B'([V_{\rm eff}]_{-})$, we get \eqref{eq:DensM} from \eqref{eq:alphaPos} and \eqref{eq:alphaNeg}. This finishes the proof of Theorem~\ref{thm:Jdens}.
\end{proof}

\section{Calculation of $J_{\rm INT}$}
\label{Jint}
The calculation of $\langle \psi, J_{\rm INT} \psi \rangle$ was carried through in \cite{fournais4,fournais8}. For convenience of the reader, we give an outline of a proof.

For a function ${\bf a} : {\mathbb R}^3 \rightarrow {\mathbb R}^3$ 
we define the mean field interaction term $D_{{\bf a}}$ as follows
\begin{align}
\label{eq:Dalpha}
D_{{\bf a}}(f,g) := \frac{1}{2} \iint \overline{f(x)} \frac{(x-y)\cdot\big({\bf a}(x)-{\bf a}(y)\big)}{|x-y|^3} g(y)\,dxdy.
\end{align}
The comparison between $\langle \psi, J_{\rm INT} \psi \rangle$ and $J^{\rm MTF}_{\rm INT}$ is contained in the next theorem.
\begin{thm}
\label{thm:Jint}~\\
Let ${\bf a_{sc}} = (a_{1,sc}, a_{2,sc},0) \in C_0^{\infty}({\mathbb R}^3)$ and define ${\bf a}(x) = \ell {\bf a_{sc}}(x/\ell)$ with $\ell$ from \eqref{eq:ell}. Let ${\bf \tilde{a}}$ be as defined previously. Furthermore, the operator $ J_{\rm INT}$ is defined by \eqref{eq:Jint} and $D_{{\bf \tilde{a}}}$ by \eqref{eq:Dalpha}.\\
Let $\lambda>0$ and $\beta_{\infty} \in [0, +\infty]$ and let $(N, Z, B) =(N_n, Z_n, B_n)$ be a sequence 
with $Z \rightarrow \infty$, and 
such that 
$$
\lambda = N/Z, \quad B/Z^{4/3} \rightarrow \beta_{\infty}, \quad B/Z^3 \rightarrow 0,
$$
(as $n \rightarrow \infty$).
Then, if $\psi= \psi_{N,Z,B}$ is an associated sequence of ground states of $H(N,Z,B)$,
\begin{align}
\frac{1}{{\mathcal E}(Z,B)}\Big|\langle \psi, J_{\rm INT} \psi \rangle - D_{{\bf \tilde{a}}}( \rho^{\rm MTF}_{B,N,Z}, \rho^{\rm MTF}_{B,N,Z})  \Big|
\rightarrow 0.
\end{align}
\end{thm}

\begin{proof}[Sketch of proof Theorem~\ref{thm:Jint}]~\\
The strategy of the proof is similar to the one of Theorem~\ref{thm:Jdens}.
We consider, for $\alpha \in [-\alpha_0, \alpha_0]$, and $\alpha_0 > 0$ sufficiently small, the self-adjoint operator 
$$
H^{\rm int}_{\alpha}(N,Z,B) := H(N,Z,B) + \alpha J_{\rm INT},
$$
and define $E^{\rm int}_{\alpha}(N,Z,B) := \inf \Spec H^{\rm int}_{\alpha}(N,Z,B)$.

When going through the steps of the proof of Theorem~\ref{thm:Jdens}---but for the new operator---the main difficulty is to make sure that analogues of \eqref{eq:DPos} and \eqref{eq:LO} hold.
More precisely, we need to be able to choose $\alpha_0$ sufficiently small  that
\begin{align}
\label{eq:positive}
D(f,f) +\alpha D_{{\bf \tilde{a}}}(f,f) \geq 0,
\end{align}
for all $|\alpha | \leq \alpha_0$ and all $f$ with $D(f,f) < \infty$.
Furthermore, we need the Lieb-Oxford type inequality in \eqref{eq:ModLO} below.

These two crucial estimates are the results of Lemma~\ref{lem:LiMod} and Lemma~\ref{lem:DalphaBnd} below. With these extra ingredients the proof follows that of Theorem~\ref{thm:Jdens} with mainly notational differences and will be omitted.
\end{proof}

\begin{lemma}
\label{lem:DalphaBnd}~\\
Let ${\bf a}$ and ${\bf a}_{sc}$ be as defined in Theorem~\ref{thm:summarize_improved} and let $D_{{\bf a}}$ be defined by \eqref{eq:Dalpha}. Then there exists a constant $C>0$ (depending only on a finite number of seminorms $\| \partial^{\alpha} {\bf a}_{sc} \|_{L^{\infty}({\mathbb R}^3)}$), such that 
\begin{align}
\label{eq:DalphaBnd}
\left| D_{{\bf a}}(f,f) \right| \leq C D(f,f),
\end{align}
for all $f \in C_0^{\infty}({\mathbb R}^3)$ with $D(f,f) < \infty$.
\end{lemma}

\begin{proof}
By scaling it suffices to prove \eqref{eq:DalphaBnd} in the case $\ell = 1$, i.e. for ${\bf a}= {\bf a}_{sc}$.
Let ${\mathcal K}$ be the operator with integral kernel 
$$
{\mathcal K}(x,y) = \frac{(x-y)\cdot\big({\bf a}(x)-{\bf a}(y)\big)}{|x-y|^3}.
$$
Recall that $(4\pi)^{-1}|x-y|^{-1}$ is the integral kernel of the operator $(-\Delta)^{-1}$ which we will denote by $p^{-2}$. Using integration by parts we therefore find
$$
4 \pi {\mathcal K} = \nabla \cdot \frac{1}{p^2} {\bf a}_{sc} - {\bf a}_{sc} \cdot  \frac{1}{p^2} \nabla.
$$
We introduce a factor of $|p|^{-1}$ on each side and get after commutation
\begin{align*}
4 \pi {\mathcal K}&=
\frac{1}{|p|} \Big\{
\nabla\cdot {\bf a}_{sc} - {\bf a}_{sc}\cdot \nabla 
+
\frac{\nabla}{|p|} \cdot [{\bf a}_{sc}, |p|] - [|p|, {\bf a}_{sc}] \cdot \frac{\nabla}{|p|}
\Big\}\frac{1}{|p|} \\
&= \frac{1}{|p|} \Big\{
(\Div\, {\bf a}_{sc})
+
\frac{\nabla}{|p|} \cdot [{\bf a}_{sc}, |p|] - [|p|, {\bf a}_{sc}] \cdot \frac{\nabla}{|p|}
\Big\}\frac{1}{|p|} .
\end{align*}
To finish the proof of Lemma~\ref{lem:DalphaBnd} we therefore only need to know that the commutator
$[|p|, \phi]$ is bounded for functions $\phi$ that are smooth and have bounded derivatives. This well-known fact can for instance be seen by splitting $|p|$ in a smooth, unbounded part---for which pseudodifferential calculus gives the result---and a compactly supported part for which the commutator is bounded as a commutator between bounded operators.
\end{proof}

We now give modified correlation inequality in the spirit of Theorem~\ref{thm:LO}.
For $\psi \in L^2({\mathbb R}^{3N})$ denote by $\rho_{\psi} \in L^1({\mathbb R}^3)$ the corresponding density
$$
\rho_{\psi}(x) := \sum_{j=1}^N \int_{{\mathbb R}^{3N}} |\psi(x^{(1)}, \ldots, x^{(N)})|^2 \delta(x-x^{(j)}) \,dx^{(1)} \cdots dx^{(N)}.
$$

\begin{lemma}[Modified Lieb-Oxford inequality]
\label{lem:LiMod}~\\
Let ${\bf a}$ and ${\bf a}_{sc}$ be as defined in Theorem~\ref{thm:summarize_improved} and let $D_{{\bf a}}$ be defined by \eqref{eq:Dalpha}. Then there exist constants $\alpha_0, C_1, C_2>0$ (depending only on a finite number of seminorms $\| \partial^{\alpha} {\bf a}_{sc} \|_{L^{\infty}({\mathbb R}^3)}$), such that for all normalised $\psi \in L^2({\mathbb R}^{3N})$ 
we have the inequality
\begin{multline}
\label{eq:ModLO}
\Big \langle \psi, \sum_{j<k} \frac{ (x^{(j)}-x^{(k)})\cdot [ (x^{(j)}-x^{(k)}) + \alpha\big({\bf a}(x^{(j)})-{\bf a}(x^{(k)})\big)]}{|x^{(j)}-x^{(k)}|^3} \, \psi \Big\rangle  \\
\geq (1-C_1 \alpha^2) D(\rho_{\psi}, \rho_{\psi}) + \alpha D_{{\bf a}}(\rho_{\psi}, \rho_{\psi}) - C_2 \int_{{\mathbb R}^3}\rho_{\psi}^{4/3}\,dx,
\end{multline}
whenever $|\alpha|\leq \alpha_0$.
\end{lemma}

\begin{proof}~\\
By scaling it suffices to prove \eqref{eq:ModLO} in the case $\ell =1$.
We choose $\alpha_0$ sufficiently small that $x \mapsto x + \alpha{\bf a}_{sc}(x)$ is invertible on ${\mathbb R}^3$ for $|\alpha|\leq \alpha_0$. Let $\phi_{\alpha}$ be the inverse, i.e. $\phi_{\alpha}(x+ \alpha {\bf a}_{sc}(x)) = x$ and define
$\Lambda_{\alpha}(x) = \det(1 + \alpha D{\bf a}_{sc}(x))$. By Taylor's formula there exists $C>0$ such that
\begin{align}
\label{eq:Taylor}
&\frac{(x-y)\cdot[x-y + \alpha\big({\bf a}_{sc}(x)-{\bf a}_{sc}(y)\big)]}{|x-y|^3} - C\frac{\alpha^2}{|x-y|}\nonumber\\
&\quad\quad\quad\quad\quad\quad\leq \frac{1}{|\phi_{\alpha}(x) - \phi_{\alpha}(y)|}\nonumber\\
&\quad\quad\quad\quad\quad\quad\leq  \frac{(x-y)\cdot[x-y + \alpha\big({\bf a}_{sc}(x)-{\bf a}_{sc}(y)\big)]}{|x-y|^3} +  C\frac{\alpha^2}{|x-y|}.
\end{align}
Therefore, we get (for sufficiently small $\alpha$),
\begin{align}
\label{eq:CVar}
\Big \langle \psi, \sum_{j<k} &\frac{ (x^{(j)}-x^{(k)})\cdot [ (x^{(j)}-x^{(k)}) + \alpha\big({\bf a}_{sc}(x^{(j)})-{\bf a}_{sc}(x^{(k)})\big)]}{|x^{(j)}-x^{(k)}|^3} \, \psi \Big\rangle\nonumber\\
&\geq (1-C\alpha^2)
\langle \psi, \sum_{j<k} \frac{1}{|\phi_{\alpha}(x^{(j)})-\phi_{\alpha}(x^{(k)})|} \psi \rangle\nonumber\\
&=
(1-C\alpha^2) \langle \psi_{\alpha} , \sum_{j<k}  \frac{1}{|x^{(j)}-x^{(k)}|} \psi_{\alpha} \rangle,
\end{align}
where (with the product being over $1\leq j \leq N$),
$$
\psi_{\alpha}(x_1,\ldots,x_N)=\sqrt{\prod \Lambda_{\alpha}(x^{(j)})}\,
\psi\big(x^{(1)}+ \alpha {\bf a}_{sc}(x^{(1)}),\ldots, x^{(N)}+ \alpha {\bf a}_{sc}(x^{(N)})\big).
$$
Now the standard Lieb-Oxford inequality, Theorem~\ref{thm:LO}, followed by \eqref{eq:Taylor} imply that (with $\rho_{\alpha}$ being the density of $\psi_{\alpha}$)
\begin{align}
\label{eq:StLO}
\langle \psi_{\alpha} , &\sum_{j<k}  \frac{1}{|x^{(j)}-x^{(k)}|} \psi_{\alpha} \rangle
\geq
D(\rho_{\alpha},\rho_{\alpha}) - C_{LO} \int_{{\mathbb R}^3} \rho_{\alpha}^{4/3}(x)\,dx \nonumber \\
&\quad\quad\quad\quad\geq
\frac{1}{2} \iint \frac{\rho(x)\rho(y)}{|\phi_{\alpha}(x)-\phi_{\alpha}(y)|} \,dxdy - C_{LO} \int_{{\mathbb R}^3} \rho_{\alpha}^{4/3}(x)\,dx\nonumber \\
&\quad\quad\quad\quad\geq
(1-C\alpha^2)D(\rho,\rho) + {\alpha} D_{{\bf a}}(\rho, \rho) -2C_{LO} \int_{{\mathbb R}^3} \rho^{4/3}(x)\,dx,
\end{align}
where the estimates are for small $\alpha$. Putting together \eqref{eq:CVar} and \eqref{eq:StLO} and using Lemma~\ref{lem:DalphaBnd} we get \eqref{eq:ModLO}.
\end{proof}

\section{Calculation of $J_{\rm KIN}$}
\label{KIN}
\subsection{The result}~\\
The analysis of $\langle \psi, J_{\rm KIN} \psi \rangle$ is very different in the two regimes $B \lesssim Z^{4/3}$ and $B\gg Z^{4/3}$. This reflects on the one hand the localisation to the lowest Landau band in the high $B$ regime, and on the other hand the vanishing of $J^{\rm MTF}_{\rm KIN}$ in the same parameter domain. The case $B \lesssim Z^{4/3}$ was already treated in \cite{fournais4}. We give the main ideas for completeness.

\begin{thm}
\label{thm:Jkin}~\\
Let ${\bf a_{sc}}=(a_{1,sc},a_{2,sc},0) \in C_0^{\infty}({\mathbb R}^3)$ and define ${\bf a}(x) = \ell {\bf a_{sc}}(x/\ell)$ with $\ell$ from \eqref{eq:ell}. Let ${\bf \tilde{a}}, M_{\bf a}, J_{\rm KIN}$ be as defined in Theorem~\ref{thm:Commutator} and let $J^{\rm MTF}_{\rm KIN}$ be as defined in \eqref{eq:J-MTF}.

Let $\lambda>0$ and $\beta_{\infty} \in [0, +\infty]$, and let $(N, Z, B) =(N_n, Z_n, B_n)$ be a sequence 
with $Z \rightarrow \infty$, and 
such that 
$$
\lambda = N/Z, \quad \quad B/Z^{4/3} \rightarrow \beta_{\infty},\quad\quad B/Z^3 \rightarrow 0,
$$
(as $n \rightarrow \infty$).
Then, if $\psi= \psi_{N,Z,B}$ is an associated sequence of ground states of $H(N,Z,B)$,
\begin{align}
\label{eq:JKinLille}
\frac{1}{ {\mathcal E}(Z,B)}\Big | \langle \psi, J_{\rm KIN} \psi \rangle - J^{\rm MTF}_{\rm KIN} \Big| \rightarrow 0.
\end{align}
In the case $\beta_{\infty} = + \infty$, \eqref{eq:JKinLille} is improved to 
\begin{align}
\label{eq:JKinStor}
\frac{1}{ {\mathcal E}(Z,B)}\Big\{ \Big | \langle \psi, J_{\rm KIN} \psi \rangle \Big| + | J^{\rm MTF}_{\rm KIN} | \Big\} \rightarrow 0.
\end{align}
\end{thm}

\subsection{Case of $\beta_{\infty} < + \infty$}~\\
Here we will prove \eqref{eq:JKinLille} in the case $B \lesssim Z^{4/3}$. We introduce 
$$
H_{\alpha}^{\rm kin}(N,Z,B):= H(N,Z,B) + \alpha J_{\rm KIN},
$$
and proceed as for the calculation of $J_{\rm DENS}$. Clearly the crucial point is to establish a semiclassical result similar to Theorem~\ref{thm:MagnSemi}. This was obtained in \cite{fournais4} from which we get

\begin{thm}\label{thm:KinSemiclass}~\\
Let $U_{B,Z}$ be a potential satisfying the hypothesis from Theorem~\ref{thm:MagnSemi}. Define
\begin{align}
\hat{P}_{B,u,t}(v) := \frac{2B}{3\pi} \sum_{\nu = 0}^{\infty}
d_{\nu} \frac{b_{u,t}}{\Lambda_{u,t}} \big[ (2\nu +1)Bb_{u,t}-B(1+2t b_3(u)) - v \big]_{-}^{3/2}, 
\end{align}
with $d_0 := \frac{1}{2\pi}$, $d_{\nu} := \pi^{-1}$ for $\nu \geq 1$.
Here 
\begin{align*}
b_{u,t} &:= \big| \curl_{x} \sqrt{1+tM_{\bf a}(u)} {\bf A}\big( \sqrt{1+tM_{\bf a}(u)} x\big) \big|
= 1 + t b_3(u) + {\mathcal O}(t^2),\\
\Lambda_{u,t} &:= | \det(\sqrt{1+tM_{\bf a}(u)})| = 1 - t b_3(u) + {\mathcal O}(t^2).
\end{align*}
Let $\lambda>0$ and $\beta_{\infty} \in [0, +\infty)$, and let $(N, Z, B) =(N_n, Z_n, B_n)$ be a sequence 
with $Z \rightarrow \infty$, and 
such that 
$$
\lambda = N/Z, \quad \quad B/Z^{4/3} \rightarrow \beta_{\infty},
$$
(as $n \rightarrow \infty$). Then
\begin{align}
- \tr[ {\mathfrak h}^1_{t}(B, U_{B,Z}]_{-} \geq - \int_{{\mathbb R}^3} \hat{P}_{B,u,t}(U_{B,Z}(u))\,du
+ o({\mathcal E}(B,Z)).
\end{align}
\end{thm}

Using Theorem~\ref{thm:KinSemiclass} instead of Theorem~\ref{thm:MagnSemi} one obtains \eqref{eq:JKinLille} for $\beta_{\infty} < + \infty$ by the same method as for the proof of Theorem~\ref{thm:Jdens}. We omit the details. Notice though, that the second order (in $t$) difference between $b_{u,t}$ and $1+tb_3$ becomes a dominant term for $B\gg Z^{4/3}$. 
On a technical level, this is the reason why we have to treat the case of large $B$ differently.

\subsection{Case of $\beta_{\infty} = +\infty$}~\\
In this region we will prove the improved estimate \eqref{eq:JKinStor}. Since, by \eqref{eq:KinMTFSmall}, $J^{\rm MTF}_{\rm KIN} = o({\mathcal E}(Z,B))$ in this regime, we only have to prove the corresponding bound on $| \langle \psi, J_{\rm KIN} \psi \rangle|$. We state and prove a slightly more general bound.

\begin{thm}
\label{thm:JKIN}~\\
Let ${\bf b}_{{\rm sc}} = (b_{{\rm sc},1}, b_{{\rm sc},2}, b_{{\rm sc},3}) \in C^2_0({\mathbb R}^3, {\mathbb R}^3)$ and define ${\bf b}(x) := {\bf b}_{{\rm sc}}(x/\ell)$ with $\ell = Z^{-1/3}(1+B/Z^{4/3})^{-2/5}$.
Let $M_{{\rm sc}} =\{M_{{\rm sc},i,j} \}_{i,j=1}^3$ be a $C^2_0$-function with values in the symmetric $3 \times 3$ matrices and such that
$$
M_{{\rm sc},3,3} = 0\quad\quad \tr \, M_{{\rm sc}} = 2 b_{{\rm sc},3}.
$$
Define $M(x):=M_{{\rm sc}}(x/\ell)$.
Let $\lambda >0$ be given. For all $\epsilon>0$, there exists $C>0$ such that if $\psi$ is a normalised ground state of $H(N,Z,B)$ where
\begin{align}
N/Z = \lambda, \quad
Z\geq C, \quad
CZ^{4/3}\leq B \leq C^{-1} Z^3,
\end{align}
then
\begin{align}
\label{eq:JKIN}
\big| \langle \psi, J_{{\rm KIN}} \psi \rangle \big| \leq \epsilon \,{\mathcal E}(Z,B),
\end{align}
where $J_{{\rm KIN}}$ is the operator introduced in \eqref{eq:Jkin} for the given $M, {\bf b}$.
\end{thm}

We will for shortness write \eqref{eq:JKIN} as $ \langle \psi, J_{{\rm KIN}} \psi \rangle = o({\mathcal E}(Z,B))$ without explicitly including the limits (large $Z,B$) and dependence on parameters from Theorem~\ref{thm:JKIN} in the notation. We will use this shorter notation in the proof below.

~\\
\noindent{\bf Proof of Theorem~\ref{thm:JKIN}}~\\
We will reduce the proof of Theorem~\ref{thm:JKIN} to the estimates on confinement, Theorem~\ref{thm:loc_improved} and Theorem~\ref{thm:BiggerConf}.

We write $M$ as
\begin{align}
\label{eq:MN}
M&:= {\mathcal M} + {\mathcal N} \nonumber \\
&=
\left(\begin{matrix} M_{11} & M_{12} & 0 \\ M_{12} & M_{22} & 0 \\ 0 & 0 & 0\end{matrix}\right) 
+
\left(\begin{matrix} 0 & 0 & N_{13} \\ 0 & 0 & N_{23} \\ N_{13} & N_{23} & 0 \end{matrix}\right) .
\end{align}
We split $J_{{\rm KIN}}$ accordingly
\begin{align}
\label{eq:J1-J2}
&J_{{\rm KIN}}= \sum_{j=1}^N (J_1^{(j)} + J_2^{(j)}), \nonumber\\
&J_1 := p_{{\bf A}} {\mathcal M} p_{{\bf A}} + B \sigma
\cdot {\bf b} - \frac{1}{2} \Delta_{\perp} b_3, \quad\quad
J_2 := p_{{\bf A}} {\mathcal N} p_{{\bf A}}.
\end{align}

The part of $J_{{\rm KIN}}$ which is hardest to estimate is $\sum J_1^{(j)}$. For this part we need very precise estimates on the confinement to the lowest Landau band.

\begin{lemma}
\label{lem:OffdiagonalNew}~\\
Let the assumptions be as in Theorem~\ref{thm:JKIN} and let $J_1$, $J_2$ be as defined in \eqref{eq:J1-J2}. Then 
\begin{align}
\label{eq:54}
\langle \psi, \sum_{j=1} ^N J_1^{(j)} \psi \rangle = o({\mathcal E}(Z,B)),\\
\label{eq:55}
\langle \psi, \sum_{j=1} ^N J_2^{(j)} \psi \rangle = o({\mathcal E}(Z,B)). 
\end{align}
\end{lemma}

Clearly Theorem~\ref{thm:JKIN} follows from Lemma~\ref{lem:OffdiagonalNew}. We prove the estimates \eqref{eq:54} and \eqref{eq:55} separately.

\begin{proof}[Proof of \eqref{eq:54}]~\\
\noindent{\bf Reduction to the non-diagonal part.}\\
Notice that the assumptions imply that $\Delta_{\perp} b_3 = \ell^{-2} {\mathcal U}(x/\ell)$ for some ${\mathcal U} \in C^0_0({\mathbb R}^3)$ and that for $B \geq Z^{4/3}$ we always have  the relation $\ell^{-2} \leq B$.

The operator $\hat{K}$ defining the Landau levels is unitarily equivalent to a harmonic oscillator. We can define the corresponding raising and lowering operators $a^*$ and $a$ by
\begin{align}
\label{eq:RaiseLow}
a := p_{{\bf A},1} - i p_{{\bf A},2}, \quad\quad
a^* := p_{{\bf A},1} + i p_{{\bf A},2}.
\end{align}
In particular, we have $a \Pi_0 = 0$.

A direct calculation, expressing the $p_{{\bf A}}$ in terms of $a,a^*$, gives
\begin{align}
\label{eq:J1-vanish}
\Pi_0 J_1 \Pi_0 = 0.
\end{align}
Here we used, among other things, the identities
\begin{align*}
&\Pi_0 p_{{\bf A}} {\mathcal M} p_{{\bf A}} \Pi_0 = \frac{B}{2} \Pi_0 \tr[{\mathcal M}] \Pi_0 + \frac{1}{4} 
 \Pi_0 (\Delta_{\perp} \tr[{\mathcal M}]) \Pi_0 \\
&\Pi_0 {\bf b} \cdot \sigma \Pi_0 = - \Pi_0b_3 \Pi_0.
\end{align*}

Furthermore, there exists a constant $C>0$ such that
$$
\pm J_1 \leq C( \hat{K} + B).
$$
Thus, 
$$
\pm \Pi_> J_1 \Pi_> \leq 2C \hat{K}.
$$
Using (a weak version of) Corollary~\ref{cor:kinetic} we therefore get
\begin{align}
\label{eq:DiagUp}
 \langle \psi, \sum_{j=1}^N \Pi_>^{(j)} J_1^{(j)} \Pi_>^{(j)} \psi \rangle = o({\mathcal E}(Z,B)).
\end{align}
Combining \eqref{eq:J1-vanish} and \eqref{eq:DiagUp} we estimate the diagonal part of $\sum J_1^{(j)}$.
Thus only the off-diagonal part,
$$
 J_{1,\rm off} := \sum_{j=1}^N \big( \Pi_0^{(j)} J_1^{(j)} \Pi_>^{(j)} +  \Pi_>^{(j)} J_1^{(j)} \Pi_0^{(j)} \big),
$$
remains to be estimated. The way we estimate $J_{1,\rm off}$ will depend on the magnitude of $B$.

$\,$\\
\noindent{\bf Case 1. $Z^{4/3} \ll B \ll Z^{13/6}$.}\\
For $Z^{4/3} \ll B \ll Z^{13/6}$ the error bound in Theorem~\ref{thm:loc_improved} is sufficient for a direct attack.

The Cauchy-Schwarz inequality and the estimate $\| a^* \Pi_0 \| \leq \sqrt{2B}$ imply, for any $\eta >0$,
$$
\pm \big( \Pi_0 J_1 \Pi_>+  \Pi_>J_1\Pi_0 \big) \leq C \eta^{-1}B  + \eta \hat{K}.
$$
Taking $\eta= \epsilon {\mathcal R}_1^{-1}$, for some $\epsilon >0$ and with ${\mathcal R}_1$ defined by
\eqref{eq:R1}, we get
\begin{align}
\label{eq:CS}
\big| \langle
\psi,& J_{1,\rm off} \psi \rangle\big| 
\leq
C \epsilon^{-1} \frac{N}{Z} {\mathcal R}_1 \big( \frac{B}{Z^{4/3}} \big)^{3/5} \Big\{ Z^{7/3}  \big( \frac{B}{Z^{4/3}} \big)^{2/5} \Big\}
+ \epsilon  {\mathcal R}_1^{-1} \langle \psi, \hat{K}^N \psi \rangle.
\end{align}
Notice that, in the parameter regime studied, we have the relations
$$
Z^{7/3} (B/Z^{4/3})^{2/5} = {\mathcal E}(Z,B), \quad\quad
{\mathcal R}_1 \ll (B/Z^{4/3})^{-3/5}.
$$
Therefore, due to Corollary~\ref{cor:kinetic}, \eqref{eq:CS} implies
\begin{align}
\label{eq:ConBSmall}
\langle \psi, J_{1,\rm off} \psi \rangle = o({\mathcal E}(Z,B)).
\end{align}
Combining \eqref{eq:ConBSmall} with \eqref{eq:J1-vanish} and \eqref{eq:DiagUp} finishes the proof of \eqref{eq:54} in the case $Z^{4/3} \ll B \ll Z^{13/6}$.

$\,$\\
\noindent{\bf Case 2. $Z^{13/6}\lesssim B$.}\\
For $Z^{13/6}\lesssim B$ we do not have a sufficiently precise estimate on the confinement, so we need to use also Theorem~\ref{thm:BiggerConf}. The analysis below is valid for $Z^2 \ll B \leq 2Z^3$. 
We decompose $J_{1,\rm off}$, as 
\begin{align}
J_{1,\rm off} = {\mathfrak J}_1+ {\mathfrak J}_2 + {\mathfrak J}_3,
\end{align}
with
$$
\tilde{p}_{\rm hf}:= 1_{\{|p_3|> \frac{1}{2} \delta^{-1} L^{-1}\}}, \quad \tilde{p}_{\rm lf}:= 1_{\{|p_3|\leq \frac{1}{2}\delta^{-1} L^{-1}\}},
$$
and (with $P_0, P_>$ and $p_{\rm hf}, p_{\rm lf}$ as in Theorem~\ref{thm:BiggerConf})
\begin{align}
{\mathfrak J}_1&:=\sum_{j=1}^N \Big(\Pi_0^{(j)} J_1^{(j)} P_>^{(j)} + P_>^{(j)} J_1^{(j)} \Pi_0^{(j)} \Big),\nonumber\\
{\mathfrak J}_2&:=\sum_{j=1}^N \Big(\tilde{p}_{\rm hf}^{(j)}\Pi_0^{(j)} J_1^{(j)} \Pi_>^{(j)}p_{\rm hf}^{(j)} + p_{\rm hf}^{(j)}\Pi_>^{(j)} J_1^{(j)} \Pi_0^{(j)} \tilde{p}_{\rm hf}^{(j)}  \Big),\nonumber\\
{\mathfrak J}_3&:=\sum_{j=1}^N \Big(\tilde{p}_{\rm lf}^{(j)}\Pi_0^{(j)} J_1^{(j)} \Pi_>^{(j)}p_{\rm hf}^{(j)} + p_{\rm hf}^{(j)}\Pi_>^{(j)} J_1^{(j)} \Pi_0^{(j)}\tilde{p}_{\rm lf}^{(j)} \Big),
\end{align}
The last component, ${\mathfrak J}_3$ satisfies the estimate
$$
\pm {\mathfrak J}_3 \leq C NB (\delta L)^{M},
$$
for all $M \in {\mathbb N}$. 
This follows by standard semiclassical pseudodifferential calculus as in \cite{Ro87}, since $\tilde{p}_{\rm lf} p_{\rm hf} = 0$, so therefore the operator in question has vanishing symbol.

We estimate ${\mathfrak J}_1$ for any $\epsilon_1>0$, using the Cauchy-Schwarz inequality and Corollary~\ref{cor:Biggerkinetic},
\begin{align}
\big|\langle \psi, {\mathfrak J}_1 \psi \rangle\big|
&\leq
C \Big( \epsilon_1^{-1} {\mathcal R}_2 N B + \epsilon_1 {\mathcal R}_2^{-1} \langle \psi, \hat{K}^N  P_>^N\psi \rangle\Big) \nonumber\\
&=
C\Big( \epsilon_1^{-1} \frac{{\mathcal R}_2 N B}{{\mathcal E}(Z,B)} + \epsilon_1 \Big) {\mathcal E}(Z,B).
\end{align}
Choosing $\epsilon_1 = \big( \frac{{\mathcal R}_2 N B}{{\mathcal E}(Z,B)}\big)^{1/2}$ with ${\mathcal R}_2$ from \eqref{eq:R2}, we get
\begin{align}
\label{eq:J1}
\big|\langle \psi, {\mathfrak J}_1 \psi \rangle\big|
\leq
C\Big(\frac{{\mathcal R}_2 N B}{{\mathcal E}(Z,B)}\Big)^{1/2}{\mathcal E}(Z,B).
\end{align}
Remember that ${\mathcal R}_2$ depends on a parameter $\delta$ which we will choose when we have 
estimated ${\mathfrak J}_2$.
For any $\epsilon_2>0$ we get, using Corollary~\ref{cor:kinetic}, 
\begin{align}
\big|\langle \psi, {\mathfrak J}_2 \psi \rangle\big|
&\leq
C \sum_{j=1}^N \langle \psi, (\epsilon_2^{-1} {\mathcal R}_1 B  \tilde{p}_{\rm hf}^{(j)} + \epsilon_2 {\mathcal R}_1^{-1} \hat{K}^{(j)})\psi\rangle\nonumber\\
&\leq
C \sum_{j=1}^N \Big\langle \psi, \big(\epsilon_2^{-1} {\mathcal R}_1 B \delta^2 L^2 (p_3^{(j)})^2 + \epsilon_2 {\mathcal R}_1^{-1} \hat{K}^{(j)}\big)\psi \Big\rangle\nonumber\\
&\leq C \big(\epsilon_2^{-1}  {\mathcal R}_1 B \delta^2 L^2 + \epsilon_2\big){\mathcal E}(Z,B).
\end{align}
We choose $\epsilon_2=\delta L\sqrt{ {\mathcal R}_1 B} $ and get 
\begin{align}
\label{eq:J2}
\big|\langle \psi, {\mathfrak J}_2 \psi \rangle\big|
&\leq C  \delta L\sqrt{ {\mathcal R}_1 B}  {\mathcal E}(Z,B).
\end{align}

Remembering that for $Z^2 \ll B \leq 2 Z^3$,
$$
L=Z^{-2/5} B^{-1/5},\quad {\mathcal R_1}= \beta^{-3/5},
\quad  {\mathcal R_2}= \delta^{-1} \frac{Z}{\sqrt{B}}\beta^{-3/5} (\delta L B^{1/2})^{\mu},
$$
we get
$$
\frac{{\mathcal R}_2 N B}{{\mathcal E}(Z,B)} = \frac{Z}{\delta \sqrt{B}} (\delta L B^{1/2})^{\mu}
\leq \frac{Z}{(\delta \sqrt{B})^{1-\mu}}, \quad\quad
\delta L\sqrt{ {\mathcal R}_1 B}= \delta.
$$
Comparing \eqref{eq:J1} and \eqref{eq:J2}, and remembering the condition
\eqref{eq:deltaLower}, we therefore have to choose $\delta\ll 1$ and $\mu<1/2$ subject to the restriction
$$
\max\big(\beta^{-3/10}, Z^{1/(1-\mu)} B^{-1/2}\big) \ll \delta.
$$
Clearly this is possible in the parameter regime $B \geq Z^{2 + 1/10}$, for instance the choice 
$$
\mu = \frac{1}{100},\quad\quad 
\delta = \Big\{ \max\big(\beta^{-3/10}, Z^{1/(1-\mu)} B^{-1/2}\big)\Big\}^{1/3},
$$ 
works.
So we have proved that for $Z^{2 + 1/10} \leq B \leq 2Z^3$, 
\begin{align}
\label{eq:ConcBBig}
\langle \psi, J_{1,\rm off} \psi \rangle = o({\mathcal E}(Z,B)).
\end{align}
Combining \eqref{eq:ConBSmall} and \eqref{eq:ConcBBig} we get \eqref{eq:54}.
\end{proof}

\begin{remark}~\\
For $B\geq 2Z^3$, we find
$$
\frac{{\mathcal R}_2 N B}{{\mathcal E}(Z,B)} = \frac{{\mathcal R}_2 B}{Z^2(\log \frac{B}{Z^3})^2},\quad\quad
\delta L\sqrt{ {\mathcal R}_1 B}= \delta \sqrt{ {\mathcal R}_1 \frac{ B}{Z^2}}.
$$
We need both these terms to be $o(1)$. We insert ${\mathcal R}_1 =B^{-1/3}$ and ${\mathcal R}_2 = \delta^{-1} \frac{Z}{\sqrt{B}} B^{-1/3}$ and find that both terms can be made $o(1)$ as long as $B \leq Z^{4-\tilde{\mu}}$ for some $\tilde{\mu} >0$.
Therefore, our results actually also permit an analysis of the current for magnetic field strengths $B$
much stronger than $Z^3$, i.e. $B \leq Z^{4-\tilde{\mu}}$.
For $Z^3 \lesssim B$, however, MTF-theory fails to correctly approximate the behaviour of the ground state energy and it is unclear what to substitute for $\frac{d}{dt} {\big |}_{t=0} E^{{\rm MTF}}(N,Z,B{\bf e}_z+t B \curl {\bf a})$ in \eqref{eq:result_current}. In \cite{LSY1} a density matrix functional is analysed, but it is not known how to correctly generalise that functional to non-constant fields.
\end{remark}

\begin{proof}[Proof of \eqref{eq:55}]~\\
Using the raising and lowering operators from \eqref{eq:RaiseLow} we find (with $N_{jk}$ being the entries of the matrix ${\mathcal N}$ in \eqref{eq:MN})
$$
\Pi_0 J_2 \Pi_0 = \Pi_0 [a, N_{13}+ iN_{23}] \Pi_0 p_3 + \big(\Pi_0 [a, N_{13}+ iN_{23}] \Pi_0 p_3\big)^*
$$
We can write the commutator as $ [a, N_{13}+ iN_{23}] = \ell^{-1} \phi(x/\ell)$ for some $\phi \in C_0({\mathbb R}^3)$. By the Cauchy-Schwarz inequality we therefore get
$$
\pm \Pi_0 J_2 \Pi_0 \leq \epsilon^{-1} \ell^{-2} \Pi_0 |\phi(x/\ell)|^2 \Pi_0
+ \epsilon p_3^2.
$$
Since $\ell^{-1} \ll Z$ for $B \ll Z^3$ we therefore find, using Proposition~\ref{prop:apriori},
$$
\Big\langle \psi, \sum_{j=1} ^N \Pi_0^{(j)} J_2^{(j)}\Pi_0^{(j)} \psi \Big\rangle = o({\mathcal E}(Z,B)).
$$

We also write
$$
\Pi_> J_2 \Pi_> = \Pi_> A \Pi_> p_3 + p_3  \Pi_>A^* \Pi_>,
$$
with $A$ being the operator 
\begin{align}
\label{eq:A}
A:= p_{{\bf A},1} N_{13} + p_{{\bf A},2} N_{23}.
\end{align}
Notice that, since the functions $N_{ij}$ are bounded,
$$
\Pi_> A \Pi_> A^* \Pi_> \leq \Pi_> A  A^* \Pi_> \leq C \hat{K}.
$$
Therefore
$$
\pm \Pi_> J_2 \Pi_> \leq \epsilon p_3^2 + \epsilon^{-1} C \hat{K},
$$
for any $\epsilon>0$, which together with Proposition~\ref{prop:apriori} and (a weak form of) Theorem~\ref{thm:loc_improved}, implies
$$
\Big\langle \psi, \sum_{j=1} ^N \Pi_>^{(j)} J_2^{(j)}\Pi_>^{(j)} \psi \Big\rangle = o({\mathcal E}(Z,B)).
$$

Finally the off-diagonal terms which we write as
\begin{align}
\label{eq:offDiag}
\Pi_0 J_2 \Pi_> + \Pi_>  J_2 \Pi_0 =
2 \big\{ \Pi_> A \Pi_0 p_3 + p_3 \Pi_0 A^* \Pi_> \big\}
+
\Pi_0 \tilde{A} \Pi_>
+
\Pi_>\tilde{A}^* \Pi_0,
\end{align}
with $A$ from \eqref{eq:A} and 
$$
\tilde{A} := p_{{\bf A},1} [N_{13},p_3] + p_{{\bf A},2} [N_{23}, p_3].
$$
We estimate, using $\| \Pi_0 a \| \leq C \sqrt{B}$,
\begin{align*}
(\Pi_0 A^* \Pi_>)^* \Pi_0 A^* \Pi_> &\leq C \hat{K}, \\
\Pi_0 \tilde{A} \Pi_>
+
\Pi_>\tilde{A}^* \Pi_0 &
\leq \epsilon^{-1} B \Pi_> + \epsilon \ell^{-2} \Pi_0 |\phi(x/\ell)|^2 \Pi_0,
\end{align*}
for some function $\phi \in C_0({\mathbb R}^3)$.
Thus, applying the Cauchy-Schwarz inequality to \eqref{eq:offDiag},
\begin{align}
\pm\big( \Pi_0 J_2 \Pi_> + \Pi_>  J_2 \Pi_0  \big) \leq
\epsilon p_3^2 + 2C  \epsilon^{-1}  \hat{K}
+ \epsilon \ell^{-2} \Pi_0 |\phi(x/\ell)|^2 \Pi_0.
\end{align}
Since $\ell^{-1} \ll Z$ for $B \ll Z^3$, we therefore conclude from Proposition~\ref{prop:apriori} and (a weak form of) Theorem~\ref{thm:loc_improved} that
$$
\Big\langle \psi, \sum_{j=1} ^N \big(\Pi_0^{(j)} J_2^{(j)}\Pi_>^{(j)} + \Pi_>^{(j)} J_2^{(j)}\Pi_0^{(j)} \big)\psi \Big\rangle = o({\mathcal E}(Z,B)).
$$
This finishes the proof of \eqref{eq:55} and therefore of Lemma~\ref{lem:OffdiagonalNew}.
\end{proof}

Clearly  Lemma~\ref{lem:OffdiagonalNew} was all that remained to be proved in order to establish Theorem~\ref{thm:JKIN}.
\qed

\appendix
\section{Estimates on $p_{\rm lf} \frac{1}{|x|} p_{\rm lf}$}
In this section we consider the operator 
$$
1_{\{|p_3| \leq \gamma\}} \frac{1}{|x|^s}1_{\{|p_3| \leq \gamma\}}
$$ 
for any $\gamma >0$ and $s\geq 1$.
We will in particular be interested in the case $s=1$.
Of course, the index `3' denotes the third component; $x= (x_1,x_2,x_3)$ and $p=(p_1,p_2,p_3)=-i\nabla$.
Informally speaking, the cut-off in frequency prohibits localisation in space on length scales shorter than $\gamma^{-1}$, which explains the main result in Lemma~\ref{lem:III} below.

Let $f \in C_0^{\infty}({\mathbb R})$ be an even function, satisfying that $f \equiv 1$ on $[-1,1]$, $\supp f \subset [-2,2]$ and define
\begin{align}
f_{\gamma}(t) = f(\gamma^{-1}t).
\end{align}

For reference, we first state without proof the following result of an elementary calculation. 

\begin{lemma}~\\
Let $f \in C_0^{\infty}({\mathbb R})$ be even.
Then the operator $f_{\gamma}(p_3)$ as an operator on $L^2({\mathbb R})$ has integral kernel
$$
K_{\gamma}(x_3,y_3)= \frac{\gamma}{2\pi} \hat{f}(\gamma(x-y)),
$$
where $\hat{f} \in {\mathcal S}({\mathbb R})$ is the Fourier transform of $f$.
\end{lemma}

\begin{lemma}
\label{lem:II}~\\
Let $f \in C_0^{\infty}({\mathbb R})$ be even. Then, for all $s \geq 1$, $q>1$ there exists $C>0$ such that for all $a, \gamma>0$ we have the estimate
\begin{align*}
\Big\| f_{\gamma}(p_3) \frac{1}{(a^2 + x_3^2)^{s/2}} f_{\gamma}(p_3) \Big\|_{{\mathcal B}(L^2({\mathbb R}_{x_3}))}
\leq C \frac{(a\gamma)^{1/q}}{a^s}.
\end{align*}
\end{lemma}

\begin{proof}
The operator has integral kernel
$$
Q_{\gamma}(x,y) = \int_{{\mathbb R}} K_{\gamma}(x,w) \frac{1}{(a^2 + w^2)^{s/2}} K_{\gamma}(w,y)\,dw.
$$
By symmetry and Schur's Lemma, we have
\begin{align*}
\| Q_{\gamma} \| &\leq
\sup_{x}  \int_{{\mathbb R}} \int_{{\mathbb R}} |K_{\gamma}(x,w)| \frac{1}{(a^2 + w^2)^{s/2}} |K_{\gamma}(w,y)|\,dwdy,
\end{align*}
so by the H\"{o}lder inequality, we get with $q^{-1} + (q')^{-1} =1$,
\begin{align*}
\| Q_{\gamma} \| &\leq
\sup_{x} \frac{\gamma^2}{4\pi^2} \int_{{\mathbb R}} \int_{{\mathbb R}} \frac{|\hat{f}(\gamma(x-w))|}{(a^2 + w^2)^{s/2}}|\hat{f}(\gamma(w-y))|\,dwdy\\
&\leq
\| \hat{f} \|_{L^1}  \frac{1}{4\pi^2}\sup_{x}\int_{{\mathbb R}} \frac{|\hat{f}(u)|}{(a^2 + (x-\frac{u}{\gamma})^2)^{s/2}}\,du\\
&\leq
\frac{1}{4\pi^2} \| \hat{f} \|_{L^1}  \| \hat{f} \|_{L^{q'}}  \frac{(a\gamma)^{1/q}}{a^s} 
\Big\{\int_{{\mathbb R}} \frac{du'}{(1 + u'^2)^{sq/2}}
\Big\}^{1/q}.
\end{align*}
\end{proof}

\begin{lemma}
\label{lem:III}~\\
Let $f \in C_0^{\infty}({\mathbb R})$ be even. 
Then, for all $s\geq 1$, $q>1$ there exists $C>0$ such that for all $\gamma>0$ and all $z \in {\mathbb R}^3$, we have the lower bound
$$
-f_{\gamma}(p_3) \frac{1}{|x-z|^s} f_{\gamma}(p_3) \geq -C \frac{\gamma^{1/q}}{|x_{\perp}-z_{\perp}|^{s-1/q}},
$$
as operators on $L^2({\mathbb R}^3)$.
\end{lemma}

\begin{proof}~\\
By implementing the unitary scaling $x \mapsto \gamma^{-1}x$ and translation on ${\mathbb R}^3$, it suffices to consider the case $\gamma =1$, $z=0$.
By Lemma~\ref{lem:II},
\begin{align*}
\Big\langle \psi,  f_1(p_3) &\frac{-1}{|x|^s} f_1(p_3) \psi \Big\rangle_{L^2({\mathbb R}^3)}\nonumber\\
&=
\int_{{\mathbb R}^2_{\perp}} \Big\langle \psi,  f_1(p_3) \frac{-1}{(|x_{\perp}|^2 + |x_3|^2)^{s/2}} f_1(p_3) \psi \Big\rangle_{L^2({\mathbb R_{x_3}})}\,dx_{\perp}\\
&\geq -C \int_{{\mathbb R}^3} |\psi(x)|^2 |x_{\perp}|^{\frac{1}{q}-s}\,dx.
\end{align*}
This finishes the proof of Lemma~\ref{lem:III}.
\end{proof}

\begin{lemma}
\label{lem:IV}~\\
Let $P_{>}$ be as defined in \eqref{eq:Pbig} and
let $c_1>0$, $\mu \in(0,1/2)$. There exists $c_0>0$ such that if 
\begin{align}
\label{eq:codnition}
\frac{Z}{\epsilon B \delta L} (\delta L B^{1/2})^{\mu}
\leq c_0,
\end{align}
then, for all $z \in {\mathbb R}^3$,
$$
P_> \Big( H_{\bf A} - c_1 \epsilon^{-1} Z \frac{1}{|\cdot-z|} \Big) P_>
\geq
\frac{1}{4}B P_>.
$$
\end{lemma}

\begin{proof}~\\
Using the inequality 
$$
\Pi_> H_{\bf A} \Pi_> \geq \frac{1}{2}( {\bf p}^2_{\bf A} + B ) \Pi_>,
$$
and the result of Lemma~\ref{lem:III}, with $q^{-1} = 1 - \mu$, we find
$$
P_> \Big( H_{\bf A} - c_1 \epsilon^{-1} Z \frac{1}{|\cdot-z|} \Big) P_>
\geq
\frac{1}{2} P_>\Big( {\bf p}^2_{\bf A} + B-2 c_1 C  \frac{Z}{\epsilon (\delta L)^{1-\mu}} \frac{1}{|\cdot-z|^{\mu}}\Big) P_>.
$$
We implement unitarily the translation by $z$ and scaling by $B^{1/2}$ and end up having to prove that
$$
{\bf p}^2_{B^{-1}\bf A} + 1-2 c_1 C  \frac{Z}{\epsilon B (\delta L)^{1-\mu}} \frac{1}{|B^{-1/2} x |^{\mu}}
\geq \frac{1}{2}.
$$
But, under condition \eqref{eq:codnition} this inequality is clearly true for $c_0$ sufficiently small.
\end{proof}

       \bibliographystyle{alpha}

\end{document}